\documentclass[twocolumn,twocolappendix]{aastex631}
\usepackage{CJK}
\usepackage[utf8]{inputenc}

\newcommand\N[1]{{\color{blue} \bf #1}}

\usepackage{soul}

\begin{document}

\title{Spectropolarimetric Evolution Reveals Dual-Axis Ejecta in the Atypical Magnetar-Powered SN~2012au}

\correspondingauthor{S. DeSoto}
\email{sabrina.desoto@du.edu}
 \newcommand{\LCO}{\affiliation{Las Cumbres Observatory, 6740 Cortona Drive, Suite 102, Goleta, CA 93117-5575, USA}}
\newcommand{\UCSB}{\affiliation{Department of Physics, University of California, Santa Barbara, CA 93106-9530, USA}}
\newcommand{\KITP}{\affiliation{Kavli Institute for Theoretical Physics, University of California, Santa Barbara, CA 93106-4030, USA}}
\newcommand{\UCD}{\affiliation{Department of Physics and Astronomy, University of California, Davis, 1 Shields Avenue, Davis, CA 95616-5270, USA}}
\newcommand{\WIS}{\affiliation{Department of Particle Physics and Astrophysics, Weizmann Institute of Science, 76100 Rehovot, Israel}}
\newcommand{\OKC}{\affiliation{Oskar Klein Centre, Department of Astronomy, Stockholm University, Albanova University Centre, SE-106 91 Stockholm, Sweden}}
\newcommand{\OAPD}{\affiliation{INAF-Osservatorio Astronomico di Padova, Vicolo dell'Osservatorio 5, I-35122 Padova, Italy}}
\newcommand{\Caltech}{\affiliation{Cahill Center for Astronomy and Astrophysics, California Institute of Technology, Mail Code 249-17, Pasadena, CA 91125, USA}}
\newcommand{\GSFC}{\affiliation{Astrophysics Science Division, NASA Goddard Space Flight Center, Mail Code 661, Greenbelt, MD 20771, USA}}
\newcommand{\UMD}{\affiliation{Joint Space-Science Institute, University of Maryland, College Park, MD 20742, USA}}
\newcommand{\UCB}{\affiliation{Department of Astronomy, University of California, Berkeley, CA 94720-3411, USA}}
\newcommand{\TTU}{\affiliation{Department of Physics, Texas Tech University, Box 41051, Lubbock, TX 79409-1051, USA}}
\newcommand{\STScI}{\affiliation{Space Telescope Science Institute, 3700 San Martin Drive, Baltimore, MD 21218-2410, USA}}
\newcommand{\UT}{\affiliation{University of Texas at Austin, 1 University Station C1400, Austin, TX 78712-0259, USA}}
\newcommand{\IoA}{\affiliation{Institute of Astronomy, University of Cambridge, Madingley Road, Cambridge CB3 0HA, UK}}
\newcommand{\QUB}{\affiliation{Astrophysics Research Centre, School of Mathematics and Physics, Queen's University Belfast, Belfast BT7 1NN, UK}}
\newcommand{\IPAC}{\affiliation{Spitzer Science Center, California Institute of Technology, Pasadena, CA 91125, USA}}
\newcommand{\JPL}{\affiliation{Jet Propulsion Laboratory, California Institute of Technology, 4800 Oak Grove Dr, Pasadena, CA 91109, USA}}
\newcommand{\Southampton}{\affiliation{Department of Physics and Astronomy, University of Southampton, Southampton SO17 1BJ, UK}}
\newcommand{\LANL}{\affiliation{Space and Remote Sensing, MS B244, Los Alamos National Laboratory, Los Alamos, NM 87545, USA}}
\newcommand{\Tsinghua}{\affiliation{Physics Department and Tsinghua Center for Astrophysics, Tsinghua University, Beijing, 100084, People's Republic of China}}
\newcommand{\NAOC}{\affiliation{National Astronomical Observatory of China, Chinese Academy of Sciences, Beijing, 100012, People's Republic of China}}
\newcommand{\Itagaki}{\affiliation{Itagaki Astronomical Observatory, Yamagata 990-2492, Japan}}
\newcommand{\Einstein}{\altaffiliation{Einstein Fellow}}
\newcommand{\Hubble}{\altaffiliation{Hubble Fellow}}
\newcommand{\CfA}{\affiliation{Center for Astrophysics \textbar{} Harvard \& Smithsonian, 60 Garden Street, Cambridge, MA 02138-1516, USA}}
\newcommand{\UA}{\affiliation{Steward Observatory, University of Arizona, 933 North Cherry Avenue, Tucson, AZ 85721-0065, USA}}
\newcommand{\MPIA}{\affiliation{Max-Planck-Institut f\"ur Astrophysik, Karl-Schwarzschild-Stra\ss{}e 1, D-85748 Garching, Germany}}
\newcommand{\DSFP}{\altaffiliation{LSSTC Data Science Fellow}}
\newcommand{\HCO}{\affiliation{Harvard College Observatory, 60 Garden Street, Cambridge, MA 02138-1516, USA}}
\newcommand{\Carnegie}{\affiliation{Observatories of the Carnegie Institute for Science, 813 Santa Barbara Street, Pasadena, CA 91101-1232, USA}}
\newcommand{\TAU}{\affiliation{School of Physics and Astronomy, Tel Aviv University, Tel Aviv 69978, Israel}}
\newcommand{\Edinburgh}{\affiliation{Institute for Astronomy, University of Edinburgh, Royal Observatory, Blackford Hill EH9 3HJ, UK}}
\newcommand{\Birmingham}{\affiliation{Birmingham Institute for Gravitational Wave Astronomy and School of Physics and Astronomy, University of Birmingham, Birmingham B15 2TT, UK}}
\newcommand{\Bath}{\affiliation{Department of Physics, University of Bath, Claverton Down, Bath BA2 7AY, UK}}
\newcommand{\CTIO}{\affiliation{Cerro Tololo Inter-American Observatory, National Optical Astronomy Observatory, Casilla 603, La Serena, Chile}}
\newcommand{\Potsdam}{\affiliation{Institut f\"ur Physik und Astronomie, Universit\"at Potsdam, Haus 28, Karl-Liebknecht-Str. 24/25, D-14476 Potsdam-Golm, Germany}}
\newcommand{\INPE}{\affiliation{Instituto Nacional de Pesquisas Espaciais, Avenida dos Astronautas 1758, 12227-010, S\~ao Jos\'e dos Campos -- SP, Brazil}}
\newcommand{\UNC}{\affiliation{Department of Physics and Astronomy, University of North Carolina, 120 East Cameron Avenue, Chapel Hill, NC 27599, USA}}
\newcommand{\Ohio}{\affiliation{Astrophysical Institute, Department of Physics and Astronomy, 251B Clippinger Lab, Ohio University, Athens, OH 45701-2942, USA}}
\newcommand{\AAS}{\affiliation{American Astronomical Society, 1667 K~Street NW, Suite 800, Washington, DC 20006-1681, USA}}
\newcommand{\MMT}{\affiliation{MMT Observatory, P.O. Box 210065, University of Arizona, Tucson, AZ 85721-0065}}
\newcommand{\Geneva}{\affiliation{ISDC, Department of Astronomy, University of Geneva, Chemin d'\'Ecogia, 16 CH-1290 Versoix, Switzerland}}
\newcommand{\IUCAA}{\affiliation{Inter-University Center for Astronomy and Astrophysics, Post Bag 4, Ganeshkhind, Pune, Maharashtra 411007, India}}
\newcommand{\CMU}{\affiliation{Department of Physics, Carnegie Mellon University, 5000 Forbes Avenue, Pittsburgh, PA 15213-3815, USA}}
\newcommand{\NAOJ}{\affiliation{Division of Science, National Astronomical Observatory of Japan, 2-21-1 Osawa, Mitaka, Tokyo 181-8588, Japan}}
\newcommand{\IfA}{\affiliation{Institute for Astronomy, University of Hawai`i, 2680 Woodlawn Drive, Honolulu, HI 96822-1839, USA}}
\newcommand{\UCSC}{\affiliation{Department of Astronomy and Astrophysics, University of California, Santa Cruz, CA 95064-1077, USA}}
\newcommand{\Purdue}{\affiliation{Department of Physics and Astronomy, Purdue University, 525 Northwestern Avenue, West Lafayette, IN 47907-2036, USA}}
\newcommand{\Princeton}{\affiliation{Department of Astrophysical Sciences, Princeton University, 4 Ivy Lane, Princeton, NJ 08540-7219, USA}}
\newcommand{\Moore}{\affiliation{Gordon and Betty Moore Foundation, 1661 Page Mill Road, Palo Alto, CA 94304-1209, USA}}
\newcommand{\Durham}{\affiliation{Department of Physics, Durham University, South Road, Durham, DH1 3LE, UK}}
\newcommand{\JHU}{\affiliation{Department of Physics and Astronomy, The Johns Hopkins University, 3400 North Charles Street, Baltimore, MD 21218, USA}}
\newcommand{\Toronto}{\affiliation{David A.\ Dunlap Department of Astronomy and Astrophysics, University of Toronto,\\ 50 St.\ George Street, Toronto, Ontario, M5S 3H4 Canada}}
\newcommand{\Duke}{\affiliation{Department of Physics, Duke University, Campus Box 90305, Durham, NC 27708, USA}}
\newcommand{\NCU}{\affiliation{Graduate Institute of Astronomy, National Central University, 300 Jhongda Road, 32001 Jhongli, Taiwan}}
\newcommand{\Columbia}{\affiliation{Department of Physics and Columbia Astrophysics Laboratory, Columbia University, Pupin Hall, New York, NY 10027, USA}}
\newcommand{\Flatiron}{\affiliation{Center for Computational Astrophysics, Flatiron Institute, 162 5th Avenue, New York, NY 10010-5902, USA}}
\newcommand{\CIERA}{\affiliation{Center for Interdisciplinary Exploration and Research in Astrophysics and Department of Physics and Astronomy, \\Northwestern University, 1800 Sherman Avenue, 8th Floor, Evanston, IL 60201, USA}}
\newcommand{\GeminiNorth}{\affiliation{Gemini Observatory, 670 North A`ohoku Place, Hilo, HI 96720-2700, USA}}
\newcommand{\Keck}{\affiliation{W.~M.~Keck Observatory, 65-1120 M\=amalahoa Highway, Kamuela, HI 96743-8431, USA}}
\newcommand{\UW}{\affiliation{Department of Astronomy, University of Washington, 3910 15th Avenue NE, Seattle, WA 98195-0002, USA}}
\newcommand{\catalyst}{\altaffiliation{LSSTC Catalyst Fellow}}
\newcommand{\USask}{\affiliation{Department of Physics \& Engineering Physics, University of Saskatchewan, 116 Science Place, Saskatoon, SK S7N 5E2, Canada}}
\newcommand{\Thacher}{\affiliation{Thacher School, 5025 Thacher Road, Ojai, CA 93023-8304, USA}}
\newcommand{\Rutgers}{\affiliation{Department of Physics and Astronomy, Rutgers, the State University of New Jersey,\\136 Frelinghuysen Road, Piscataway, NJ 08854-8019, USA}}
\newcommand{\FSU}{\affiliation{Department of Physics, Florida State University, 77 Chieftan Way, Tallahassee, FL 32306-4350, USA}}
\newcommand{\Melbourne}{\affiliation{School of Physics, The University of Melbourne, Parkville, VIC 3010, Australia}}
\newcommand{\ASTROthreeD}{\affiliation{ARC Centre of Excellence for All Sky Astrophysics in 3 Dimensions (ASTRO 3D)}}
\newcommand{\Stromlo}{\affiliation{Mt.\ Stromlo Observatory, The Research School of Astronomy and Astrophysics, Australian National University, ACT 2601, Australia}}
\newcommand{\NCPAS}{\affiliation{National Centre for the Public Awareness of Science, Australian National University, Canberra, ACT 2611, Australia}}
\newcommand{\TAMU}{\affiliation{Department of Physics and Astronomy, Texas A\&M University, 4242 TAMU, College Station, TX 77843, USA}}
\newcommand{\Mitchell}{\affiliation{George P.\ and Cynthia Woods Mitchell Institute for Fundamental Physics \& Astronomy, College Station, TX 77843, USA}}
\newcommand{\ESO}{\affiliation{European Southern Observatory, Alonso de C\'ordova 3107, Casilla 19, Santiago, Chile}}
\newcommand{\ICE}{\affiliation{Institute of Space Sciences (ICE, CSIC), Campus UAB, Carrer
de Can Magrans, s/n, E-08193 Barcelona, Spain}}
\newcommand{\IEEC}{\affiliation{Institut d'Estudis Espacials de Catalunya, Gran Capit\`a, 2-4, Edifici Nexus, Desp.\ 201, E-08034 Barcelona, Spain}}
\newcommand{\Warwick}{\affiliation{Department of Physics, University of Warwick, Gibbet Hill Road, Coventry CV4 7AL, UK}}
\newcommand{\Macquarie}{\affiliation{School of Mathematical and Physical Sciences, Macquarie University, NSW 2109, Australia}}
\newcommand{\AAARC}{\affiliation{Astronomy, Astrophysics and Astrophotonics Research Centre, Macquarie University, Sydney, NSW 2109, Australia}}
\newcommand{\Capodimonte}{\affiliation{INAF - Capodimonte Astronomical Observatory, Salita Moiariello 16, I-80131 Napoli, Italy}}
\newcommand{\INFNNapoli}{\affiliation{INFN - Napoli, Strada Comunale Cinthia, I-80126 Napoli, Italy}}
\newcommand{\ICRANet}{\affiliation{ICRANet, Piazza della Repubblica 10, I-65122 Pescara, Italy}}
\newcommand{\MSU}{\affiliation{Center for Data Intensive and Time Domain Astronomy, Department of Physics and Astronomy,\\Michigan State University, East Lansing, MI 48824, USA}}
\newcommand{\SETI}{\affiliation{SETI Institute,
339 Bernardo Ave, Suite 200, Mountain View, CA 94043, USA}}
\newcommand{\IAIFI}{\affiliation{The NSF AI Institute for Artificial Intelligence and Fundamental Interactions}}
\newcommand{\ANUC}{\affiliation{Department of Astronomy, AlbaNova University Center, Stockholm University, SE-10691 Stockholm, Sweden}}

\newcommand{\Konkoly}{\affiliation{Konkoly Observatory,  CSFK, Konkoly-Thege M. \'ut 15-17, Budapest, 1121, Hungary}}
\newcommand{\ELTE}{\affiliation{ELTE E\"otv\"os Lor\'and University, Institute of Physics, P\'azm\'any P\'eter s\'et\'any 1/A, Budapest, 1117 Hungary}}
\newcommand{\SZTE}{\affiliation{Department of Experimental Physics, University of Szeged, D\'om t\'er 9, Szeged, 6720, Hungary}}
\newcommand{\IdAlta}{\affiliation{Instituto de Alta Investigaci\'on, Sede Esmeralda, Universidad de Tarapac\'a, Av. Luis Emilio Recabarren 2477, Iquique, Chile}}
\newcommand{\Kavli}{\affiliation{Kavli Institute for Cosmological Physics, University of Chicago, Chicago, IL 60637, USA}}
\newcommand{\UofChicago}{\affiliation{Department of Astronomy and Astrophysics, University of Chicago, Chicago, IL 60637, USA}}
\newcommand{\Fermi}{\affiliation{Fermi National Accelerator Laboratory, P.O.\ Box 500, Batavia, IL 60510, USA}}
\newcommand{\Dartmouth}{\affiliation{Department of Physics and Astronomy, Dartmouth College, Hanover, NH 03755, USA}}
\newcommand{\Surrey}{\affiliation{Department of Physics, University of Surrey, Guildford GU2 7XH, UK}}
\newcommand{\NU}{\affiliation{Center for Interdisciplinary Exploration and Research in Astrophysics (CIERA) and Department of Physics and Astronomy, Northwestern University, Evanston, IL 60208, USA}}

\newcommand{\itagaki}{\affiliation{Itagaki Astronomical Observatory, Yamagata 990-2492, Japan}}

\newcommand{\DU}{\affiliation{Department of Physics \& Astronomy, University of Denver, 2112 East Wesley Avenue, Denver, CO 80208, USA}}

\newcommand{\SDS}{\affiliation{Department of Astronomy, San Diego State University, San Diego, CA 92812, USA}}
\newcommand{\ARI}{\affiliation{Astrophysics Research Institute, Liverpool John Moores University, 146 Brownlow Hill, Liverpool L3 5RF, UK}}
\newcommand{\UofSh}{\affiliation{Department of Physics and Astronomy, University of Sheffield, Hicks Building, Hounsfield Road, Sheffield S3 7RH, UK}}
\newcommand{\UofH}{\affiliation{Centre for Astrophysics Research, University of Hertfordshire, Hatfield, AL10 9AB, UK}}
\newcommand{\UVA}{\affiliation{Department of Astronomy, University of Virginia, Charlottesville, VA 22904, USA}}
 \author[0000-0003-4829-6499]{Sabrina DeSoto}
 \DU
 \author[0000-0003-1495-2275]{Jennifer L. Hoffman}
 \DU
 \author[0000-0002-3452-0560]{G. Grant Williams}
 \MMT
 \UA
 \author[0000-0002-8826-3571]{Christopher Bilinski}
 \UA
 \author[0000-0001-7839-1986]{Douglas C. Leonard}
 \SDS
 \author[0000-0002-0370-157X]{Peter A. Milne}
 \UA
 \author[0000-0002-3107-8551]{Christopher Pickens}
 \DU
 \author[0000-0002-4022-1874]{Manisha Shrestha}
 \UA
 \author[0000-0001-5510-2424]{Nathan Smith}
 \UA
 \author[0000-0002-5083-3663]{Paul S. Smith}
 \UA



\begin{abstract}

We present six epochs of optical spectropolarimetric observations of the unique and slow-evolving Type Ib supernova (SN) 2012au, between 0 and 295 days post $R$-band maximum. The polarization levels seen throughout our observations are on average $0.87\% \pm 0.05\%$ higher than those of any Type Ib SN~yet studied, suggesting either that it is the most asymmetric of the sample, or if all SNe Ib have similar asymmetry, that it is viewed at a more optimum angle. Significant continuum polarization indicates that the photosphere exhibited a global departure from spherical symmetry at the level of $10\%-40\%$ at the earliest times (days 0--40), which decreased to $0\%-20\%$ by days 57--90. During the early photospheric phase, the ejecta maintained a near-constant orientation of $12\degr-20\degr$ on the sky, as shown by the dominant axis in the Stokes $q-u$ plane. Polarization signatures in the Fe \textsc{ii} $\lambda$$\lambda$$\lambda$4924, 5018, 5169 lines shared this axis. Meanwhile, high levels of polarization associated with the He \textsc{i} lines traced distinct $q-u$ loops with a dramatic rotation away from the dominant axis, indicating that the early-time ejecta were also characterized by hot, fast, helium-rich material concentrated near the poles. At day 295, during the transition to the nebular phase, a new, highly elongated structure became prominent in the ejecta, with an axis  orthogonal to the dominant axis that defined the photospheric phase. This dual-axis geometry may link SN 2012au's high luminosity and asymmetric structure to a magnetar powering mechanism.

\end{abstract}



\section{Introduction} \label{sec:intro}

 The geometrical structure of a supernova (SN) explosion contains important links to both the explosion mechanism and the stellar structure of its progenitor. However, because most supernovae (SNe) remain unresolvable point sources throughout their evolution (with the notable exception of SN 1987A, e.g. \citealt{Matsuura24}), we cannot use direct imaging to reveal their explosion geometries. Instead, polarimetric studies offer insights into SN structure by leveraging the geometrical information encoded in the scattering mechanism. Observations of nonzero continuum polarization have shown that the majority of SN explosions are aspherical \citep{Shapiro82, McCall84, Hoflich91, WW08}. Furthermore, spectropolarimetric analysis of line profiles provides valuable information about the distribution of elements in the SN, the density of interacting areas, and the presence of dust \citep[e.g.,][]{Reilly16, Tanaka08, Bilinski18, Bilinski20}. 
 

Core-collapse SNe of Type Ib are defined by a lack of hydrogen in their spectra, and are closely related to SNe Type Ic, which also lack helium \citep{Filippenko97}. Because their progenitors possess no significant H (or H and He) envelopes, these types are often referred to collectively as ``stripped-envelope supernovae’’ (SESNe).  The majority of SESNe are thought to arise from lower-mass (12--20 $M_{\odot}$) progenitors that have had their envelopes removed through mass transfer or mass loss in a binary system, rather than high-mass single stars \citep{smith11,Drout2011,lyman16}. Spectropolarimetric observations have the potential to confirm this progenitor connection by identifying explosion asymmetries and circumstellar material configurations that may arise from binary interactions. 

However, there are currently only four SNe Type Ib with published and analyzed spectropolarimetry (SP): SN~2005bf \defcitealias{Maund07}{M07} \citep{Maund07,Tanaka09}, \defcitealias{Tanaka09}{T09} SN~2008d \defcitealias{Maund09}{M09} \citep{Maund09}, 2009jf \defcitealias{Tanaka12}{T12}\citep{Tanaka12} and SN~iPTF 13bvn \defcitealias{Reilly16}{R16}\citep{Reilly16} (hereafter referred to as  \citetalias{Maund07}, \citetalias{Tanaka09}, \citetalias{Maund09}, \citetalias{Tanaka12}, and \citetalias{Reilly16}, respectively). All five papers reported significant polarization values for key line features such as Fe \textsc{ii}, He \textsc{i}, and O \textsc{i}. \citetalias{Maund07}, \citetalias{Maund09}, and \citetalias{Reilly16} found continuum polarization implying $\sim10\%$ photosphere ellipticity and suggested it was due to a stalled jet unable to break through the core of the progenitor. Several SP studies of SNe Type IIb, which differ from Type Ib only by a residual amount of surface hydrogen, have also shown similar levels of continuum and line polarization (disregarding the hydrogen features) to the Type Ib SP sample \citep{Mauerhan15, Stevance20}. In order to draw broad conclusions about these possible commonalities in SN Ib explosions, their potential connection to SNe IIb, and their implications for the progenitor systems, it is necessary to expand the limited data set of spectropolarimetric studies of SNe Ib.

A promising candidate to help further our understanding of SESNe, the unusual Type Ib SN 2012au, was discovered on March 14, 2012 \citep{Discovertele, discovertele2}. Located in the host galaxy NGC 4790, a distance of 22.917 $\pm$ 1.2 Mpc away \citep{NED}, SN~2012au has been the focus of several studies conducted in the optical and radio wavelength regimes to examine its light curve and spectral features. 
Its brightness peaked at $M_R = -18.7 \pm 0.2$ \citep{MiliD13}, in between typical Type Ibc peak magnitudes and the cutoff for a superluminous SN \citep{Drout2011, Quimby2013}. \citet{MiliD13} also identified SN~2012au as a rare hypernova which released an order of magnitude more kinetic energy ($10^{52}$ erg) than similar SNe Ibc ($10^{51}$ erg). 
\citet{Kamble14} determined from radio observations that while the shock-wave energy of SN~2012au was intermediate between those of two SNe Type Ic-bl, SN~1998bw 
and SN~2002ap \citep{Galama_1998, Iwamoto_1998, Kulkarni_1998}, SN~2012au's progenitor had a much higher mass-loss rate and metallicity than either one. These extreme attributes of SN~2012au, combined with the progenitor star mass of $\le80~M_\odot$ \citep{MiliD13}, contradict prior assumptions that very massive stars could not create explosions with this energy because their massive cores are susceptible to greater neutrino cooling, inhibiting their ability to build up energy \citep{burrows98, Fryer98, Kamble14}. However, the high-velocity spectral features and high levels of broadband polarization observed in SN 2012au suggest that the explosion was not symmetric, and that its immense explosion energy and unique brightness may have been powered by jets \citep{MiliD13, Kamble14, Pandey21}.


We carried out a more detailed exploration of SN 2012au’s explosion geometry based on observations obtained as part of the Supernova Spectropolarimetry Project \citep[SNSPOL;][]{Hoffman17,Williams18}. We here present six epochs of SP for SN~2012au from 0 to 295 days post $R$-band maximum brightness. By observing variations in the polarization and flux spectra of SN~2012au over time, we infer possible scenarios for its explosion geometry that hint at the underlying mechanism powering this unique SN. 

In this work, we compare our proposed early-time geometry of SN~2012au with findings from \citetalias{Maund07}, \citetalias{Maund09}, \citetalias{Tanaka12}, \citetalias{Reilly16}, and connect this picture to the geometry predicted by transitional-nebular observations of the object \citep{MiliD13, MiliD18}. The paper is structured as follows: Section 2 discusses observations and data reduction; details of the spectral evolution can be found in Section 3. Polarization signatures are the focus of Sections 4—6: we discuss interstellar polarization (ISP) estimates in Section 4, continuum polarization measurements in Section 5, and line polarization features in Section 6. In Section 7 we compare SN~2012au to the SNe studied in the papers listed above and develop a geometrical interpretation. We present our conclusions in Section 8.

\section{Observations and Data Reduction}\label{sec:Obs&reduction}

SN~2012au reached peak brightness in the $B$ band ($M_B = -18.1$)  on 2012 March 20 and in the $V$ ($M_V = -18.6 \pm 0.2$) and $R$ bands ($M_R = -18.7 \pm 0.2$) on 2012 March 21 ($\pm1$ day; \citealt{Takaki, MiliD13, Pandey21}). Our six epochs of spectropolarimetric observations were acquired between 0 and 295 days post $R$-band peak brightness (Table \ref{Tab:observations}). These observations were taken using the CCD Imaging/Spectropolarimeter (SPOL) on the 61'' Kuiper telescope, 90'' Bok telescope and 6.5 m MMT at the University of Arizona as part of the Supernova Spectropolarimetry (SNSPOL) project. We used a slit size of 4.1$''$ at the Kuiper and Bok telescopes and 1.5$''$ at the MMT; this returns a spectral resolution of approximately 26 \AA~for Kuiper and Bok and 16 \AA~for the MMT. A more detailed description of the SPOL instrument and telescope specification can be found in \citet{Bilinski24}. We define our epochs by continuous, multi-day observing runs where SPOL was mounted on the same telescope and denote them hereafter with the range of days post $R$-band maximum. Standard data reduction techniques were applied to combine each group of observations into a single data set, as described by \cite{Bilinski18}.

Upon observation, we chose initial bin sizes to match the intrinsic dispersion (at half-integer values) returned by the spectrograph; for Kuiper and Bok this is approximately 4 \AA/pixel and for MMT, $\approx2.5$ \AA/pixel. For our analysis we then verified that significant polarization signatures were those present in the spectra both at the initial bin size and spectral resolution of the telescope. For increased signal-to-noise ratio and a consistent display of the data, we then rebinned all polarization spectra in this paper to 15~\AA~ or 20~\AA~ (noting which was used in each figure). We also corrected for the host galaxy redshift of $z = 0.004483$ \citep{MiliD18} and trimmed the blue end of the polarization data to 4000~\AA~to remove noisy regions corresponding to detector edges. Instrumental polarization for SPOL was $< 0.1\%$ for each epoch \citep{Bilinski20, Shrestha24}. The flux spectra in this paper are presented in their original binning resolution and not strictly spectrophotometically calibrated due to slit losses and potential transparency issues (such as clouds). However, this level of calibration is sufficient, since our main interest in the flux spectra is for line identification. This data can be accessed via an accompanying online repository: \dataset[10.5281/zenodo.16950678]{} \citep{Flux_data_Zenodo}.

The traditional calculation of total polarization $p_{trad}$ (per cent) employs the Pythagorean Theorem with the measured linear Stokes $Q$ and $U$ fluxes normalized by the total intensity $I$:
\begin{equation}
    q = Q/I \qquad u = U/I
\end{equation}
\begin{equation}
    p = \sqrt{q^2 + u^2} \qquad
    \sigma_{p} = \frac{1}{p^2} \sqrt{q^2\sigma_{q}^2 + u^2\sigma_{u}^2}
\end{equation}

\noindent However, to correct for the positive definite nature of this classical formulation, we used the expression for debiased polarization presented by \citet{WardleKronberg74}. This definition, which we have adopted in previous SNSPOL publications \citep[e.g.,][]{Leonard01,Leonard21}, takes into account the uncertainties associated with low polarization magnitudes: 

\begin{equation}\label{eq:debiP2}
    p_{debi} = \pm\sqrt{|p^2 - \sigma_{p}^2|}
\end{equation}

\noindent Here the sign is chosen to match that of the quantity inside the absolute value. We compared polarization values from the classical calculation to those calculated using the debiased method but did not find a significant difference in the results. In the remainder of this paper, all total polarization values refer to $p_{debi}$ unless otherwise noted. We then calculated the polarization position angle (PA) and its uncertainty in degrees using 

\begin{equation}
    \theta = \frac{90^\circ}{\pi} \textrm{arctan}\left(\frac{u}{q}\right)
\end{equation}

\begin{equation}\label{eq:sigP}
    \sigma_{\theta} = \frac{90^\circ}{\pi}\cdot\frac{1}{p_{debi}^2} \sqrt{q^2\sigma_{u}^2 + u^2\sigma_{q}^2} 
\end{equation}

Any further mathematical assessment of the data was applied to the $Q$, $U$, and total fluxes first. We then divided out the flux and combined values as discussed above to obtain $p$, $\theta$, and associated uncertainties. When averaging was necessary (e.g., when calculating the continuum), we used error-weighted means of $Q$ and $U$ along with the associated uncertainties. We report the calculated uncertainties unless they are smaller than our adopted systematic uncertainties, which our experience of the SPOL instrument indicates is $0.05\%$ \citep{Shrestha24}. 

\begin{table}
 \centering
 \begin{tabular}{ccccc}
    \hline
    Date & Days post & UT Start &  $Q/U$  & Airmass\\
    (UT) & \textit{R}-band max  & time & exp. (s) &\\
    \hline
    \hline
    \multicolumn{5}{c}{Days 0--7 : 90" Bok\tablenotemark{b}\tablenotemark{c}\tablenotemark{d}\tablenotemark{f}}\\
    \hline
    2012 Mar 21 & 0 & 9:17:39 & 480 & 1.39\\
    2012 Mar 21 & 0 & 9:36:18 & 480 & 1.42\\ 
    2012 Mar 22 & 1 & 8:19:59 & 480 & 1.35\\
    2012 Mar 22 & 1 & 8:39:05 & 480 & 1.35\\
    2012 Mar 23 & 2 & 8:28:00 & 720 & 1.35\\
    2012 Mar 24 & 3 & 9:36:57 & 720 & 1.46\\
    2012 Mar 25 & 4 & 9:26:28 & 720 & 1.44\\
    2012 Mar 28 & 7 & 9:06:12 & 480 & 1.42\\
    2012 Mar 28 & 7 & 9:25:04 & 480 & 1.47\\
    \hline
    \multicolumn{5}{c}{Day 26: 6.5m MMT \tablenotemark{b} \tablenotemark{c} \tablenotemark{d} \tablenotemark{f}}\\
    \hline
    2012 Apr 16 & 26 & 7:09:31 & 960 & 1.36\\
    2012 Apr 16 & 26 & 7:45:33 & 960 & 1.41\\
    2012 Apr 16 & 26 & 8:21:02 & 960 & 1.51\\
    \hline
    \multicolumn{5}{c}{Days 35--40: 61" Kuiper\tablenotemark{b}\tablenotemark{f}\tablenotemark{g}\tablenotemark{h}}\\
    \hline
    2012 Apr 25 & 35 & 7:27:59 & 960 & 1.47\\
    2012 Apr 27 & 37 & 7:58:05 & 960 & 1.61\\
    2012 Apr 28 & 38 & 7:42:32 & 960 & 1.56\\
    2012 Apr 29 & 39 & 7:44:17 & 960 & 1.58\\
    2012 Apr 30 & 40 & 7:21:55 & 960 & 1.51\\
    \hline
    \multicolumn{5}{c}{Days 57--67: 61" Kuiper\tablenotemark{c}\tablenotemark{d}\tablenotemark{g}}\\
    \hline
    2012 May 17 & 57 & 5:52:28 & 960 & 1.45\\
    2012 May 24 & 64 & 4:50:53 & 960 & 1.38\\
    2012 May 27 & 67 & 5:32:26 & 960 & 1.5\\
    \hline
    \multicolumn{5}{c}{Days 85--90: 61" Kuiper\tablenotemark{c}\tablenotemark{d}\tablenotemark{g}}\\
    \hline
    2012 Jun 14 & 85 & 5:13:54 & 800 & 1.75\\
    2012 Jun 16 & 87 & 5:32:44 & 800 & 1.97\\
    2012 Jun 19 & 90 & 4:37:38 & 960 & 1.65\\
    \hline
    \multicolumn{5}{c}{Day 295: 6.5m MMT\tablenotemark{c}\tablenotemark{e}\tablenotemark{f}\tablenotemark{g}\tablenotemark{i}}\\
    \hline
    2013 Jan 10 & 295 & 11:02:57 & 960 & 1.57\\
    \hline
 \end{tabular}
 \caption{SPOL Observations of SN~2012au}
 \tablenotetext{b}{Polarized Standard Star: HD 245310}
 \tablenotetext{c}{Polarized Standard Star: Hiltner 960}
 \tablenotetext{d}{Polarized Standard Star: VI Cyg 12}
 \tablenotetext{e}{Polarized Standard Star: BD +64 106}
 \tablenotetext{f}{Flux Standard Star: G191B2B}
 \tablenotetext{g}{Flux Standard Star: BD +28 4211}
 \tablenotetext{h}{Flux Standard Star: Feige 34}
 \tablenotetext{i}{Flux Standard Star: HZ 44}
 \label{Tab:observations}
\end{table}

\section{Spectral Evolution}\label{sec: spectral_evolution}

\begin{figure}
    \centering
    \includegraphics[width=\columnwidth]{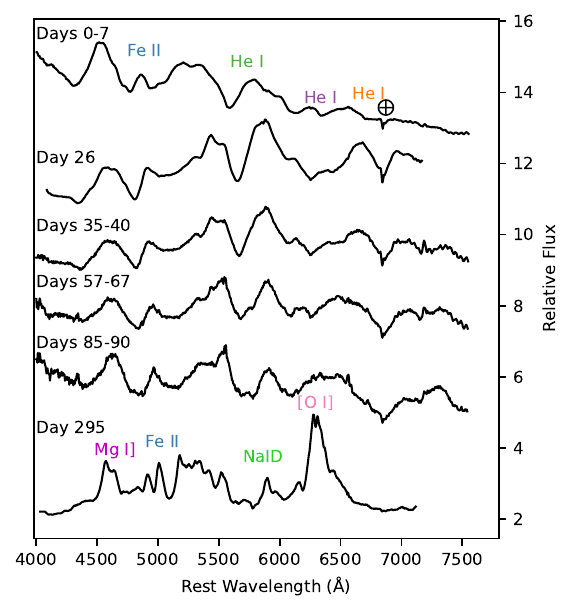}
    \caption{SN 2012au flux spectra over days 0--295 post $R$-band max, collected by the SNSPOL project. Flux spectra have been adjusted after flux calibration to stack neatly and used to identify key line features (labeled).}
    \label{fig:flux spectra}
\end{figure}

We display optical spectroscopic data for days 0--295 post $R$-band peak brightness in Figure \ref{fig:flux spectra} and in comparison to the polarized spectra in Figure \ref{fig:specpol}. Optical spectra were previously published for days -6 to +138 (post $R$-band max) by \citet{Takaki}, days -4 to +57, +274 (post $V$-band max)
by \citet{MiliD13}, -5 to +395 (post $B$-band max) by \citet{Pandey21}, and +2270 (post $V$-band max) by \citet{MiliD18}. We adopt the $R$- and $V$-band maximum light date as our reference point. Because our observations are interspersed in time with those previously presented, we were able to make reliable identifications of the spectral features by referring to these previous studies, especially \citet{Pandey21}, which used synthetic spectrum matching. The evolution of major emission and absorption features of SN~2012au can clearly be traced from days 0--90, while more narrow lines at day 295 distinguish this transitional-nebular phase from the earlier epochs.

The spectra of SN~2012au appear to be typical compared to other well-studied SNe Ib \citep{Pandey21}. The photospheric-phase spectra are each dominated by a large P Cygni profile coinciding with He \textsc{i} $\lambda$5876, and the earlier spectra also contain a secondary P Cygni profile attributed to the Fe \textsc{ii} triplet $\lambda$$\lambda$$\lambda$4924, 5018, 5169. 

\subsection{Helium Signatures and Photospheric-Phase Velocities}\label{subsec:Helium_flux_spec}
Between days 0--40, the flux spectra are each dominated by a broad P Cygni profile spanning approximately $5500-6000$~\AA~(Fig.~\ref{fig:flux spectra}). This signature corresponds to the He \textsc{i} $\lambda$5876 line commonly seen in SNe Ib. Although the Na \textsc{i} D emission line is superimposed on the peak of the P Cygni profile, the flat top of the emission component in the days 0--40 spectra suggests there is little sodium contamination \citep{Branch02}. \cite{Takaki} found the velocity of this absorption feature to be about -15000 km s$^{-1}$ around maximum light. In our data, the differences between the flux minimum of the absorption trough and the rest wavelength of the He \textsc{i} $\lambda$5876 line correspond to velocity shifts 
of -14590 ($\pm 110$) km s$^{-1}$, -11530 ($\pm 70$) km s$^{-1}$, and -10510 ($\pm 110$) km s$^{-1}$ for days 0--7, 26, and 35--40, respectively. At days 57--67, this absorption component develops a secondary dip in the flux that continues to grow in depth to day 295. The velocity shifts of the flux minima for the main absorption feature in these epochs are -10000 ($\pm 370$) km s$^{-1}$ for days 57--67 and -10510 ($\pm 370$) km s$^{-1}$ for days 85--90.



The velocities we calculated for the He \textsc{i} $\lambda$5876 and $\lambda$6678 absorption features are very similar to one another within each epoch for days 0--40 and show a slowing trend over time.
Assuming a homologous expansion, this decay in velocity over time implies that the expansion rate of the ejecta is slowing. However, it is important to distinguish the velocity of the ejecta material from the velocity of the photosphere, which we discuss in \S~\ref{subsec: iron flx}. After day 40, the absorption feature corresponding to He \textsc{i} $\lambda$6678 becomes unclear in our spectra, while a secondary absorption feature for He \textsc{i} $\lambda$5876 emerges. 
The commonality between the velocities of the shifted features suggests that a single emission region gave rise to these He \textsc{i} lines in the photospheric phase.


\subsection{Iron Multiplet}\label{subsec: iron flx} 
The Fe \textsc{ii} multiplet 42 lines ($\lambda$$\lambda$$\lambda$4924, 5018, 5169) appear as a single P Cygni profile in all our flux spectra; however, this could be blended with  He \textsc{i} $\lambda$5016. In order to analyze this spectral feature without the assumption that it is either iron or helium, we adopt 5018 \AA~as our Fe \textsc{ii} rest wavelength, as it is closest to the helium rest wavelength, resulting in similar velocity shifts for either choice of line identification. The absorption minimum of this feature is blueshifted from rest by -15410 km s$^{-1}$, -12430 km s$^{-1}$, -11830 km s$^{-1}$, and -10630 km s$^{-1}$ for days 0--7, 26, 35--40, and 57--67, respectively. We use these velocity shifts as tracers of the photospheric velocity, as was done in \citet{Branch02}. Although 
the velocities we derive for Fe \textsc{ii} in these early epochs are faster than those of helium (\S~\ref{subsec:Helium_flux_spec}), the blended nature of these lines in our spectra makes their true velocities difficult to determine.

\subsection{Oxygen} \label{subsec: Oxygen_flux}
As early as days 57--67, our spectra show a possible feature associated with the forbidden O \textsc{i} doublet [$\lambda$$\lambda$6300, 6364], which appears as a double-peaked feature at $\approx$6100 \AA. However, neither \citet{MiliD13} or \citet{Pandey21} identifies [O \textsc{i}] in their spectra until later dates closer to our observations for days 85--90. At this latest photospheric epoch, two individual peaks appear, located closer to the [O \textsc{i}] rest wavelengths than the potential feature at days 57--67.

\subsection{Transitional-Nebular Phase Spectral Signatures}\label{subsec:nebular_flux}
Our latest observation at day 295 exhibits several clear emission lines that are less pronounced or undetectable in our earlier spectra. These data also look very similar to the day 274 and 321 spectra presented by \citet{MiliD13}. At this time, individual emission features for the Fe \textsc{ii} multiplet 42 lines ($\lambda$$\lambda$$\lambda$4924, 5018, 5169) are present very close to their rest wavelengths (Fig.~\ref{fig:flux spectra}). We also see a clear distinction between the Na \textsc{i} D $\lambda\lambda$5890, 5896 features at day 295 (which appear blended with the  He \textsc{i} $\lambda$5876 P Cygni profile in our earlier epochs). At days 85--90 and 295, we identify the same [O \textsc{i}] $\lambda\lambda$6300, 6364 doublet signature and Mg \textsc{i}] $\lambda$4571 asymmetric emission feature as seen in \citet{MiliD13}. These emission lines are also common for SNe Ib \citep{Pandey21}.

The two peaks of the [O \textsc{i}] doublet are blueshifted by -1790 $\pm 360$ km s$^{-1}$ (measured from the minimum between the two peaks to the line center of $\lambda$6332), rather than appearing as a symmetric profile about their line center. As suggested by \citet{Taubenberger2009}, we compared the line profiles of Mg \textsc{i}] $\lambda$4571 and [O\textsc{i}] $\lambda\lambda$6300,6364 and found that the flux peaks for both lines are located at similar velocities with respect to the line centers ($v_{[O\textsc{i}]} = -$2610 $\pm$ 360 km s$^{-1}$ and $v_{Mg\textsc{i}]} =  -$2920 $\pm$ 490 km s$^{-1}$); this is consistent with the majority of SNe Ibc transitional-nebular spectra \citep{Taubenberger2009}. The emergence of the Mg \textsc{i}] line at this late epoch as a fairly broad, seemingly asymmetric double peak profile, could indicate that a blending with iron lines masked this feature at early epochs and may be contributing to the secondary peak of the line profile at this time \citep{Taubenberger2009}.




\begin{figure*}
    \centering
    \includegraphics[width=.7\textwidth]{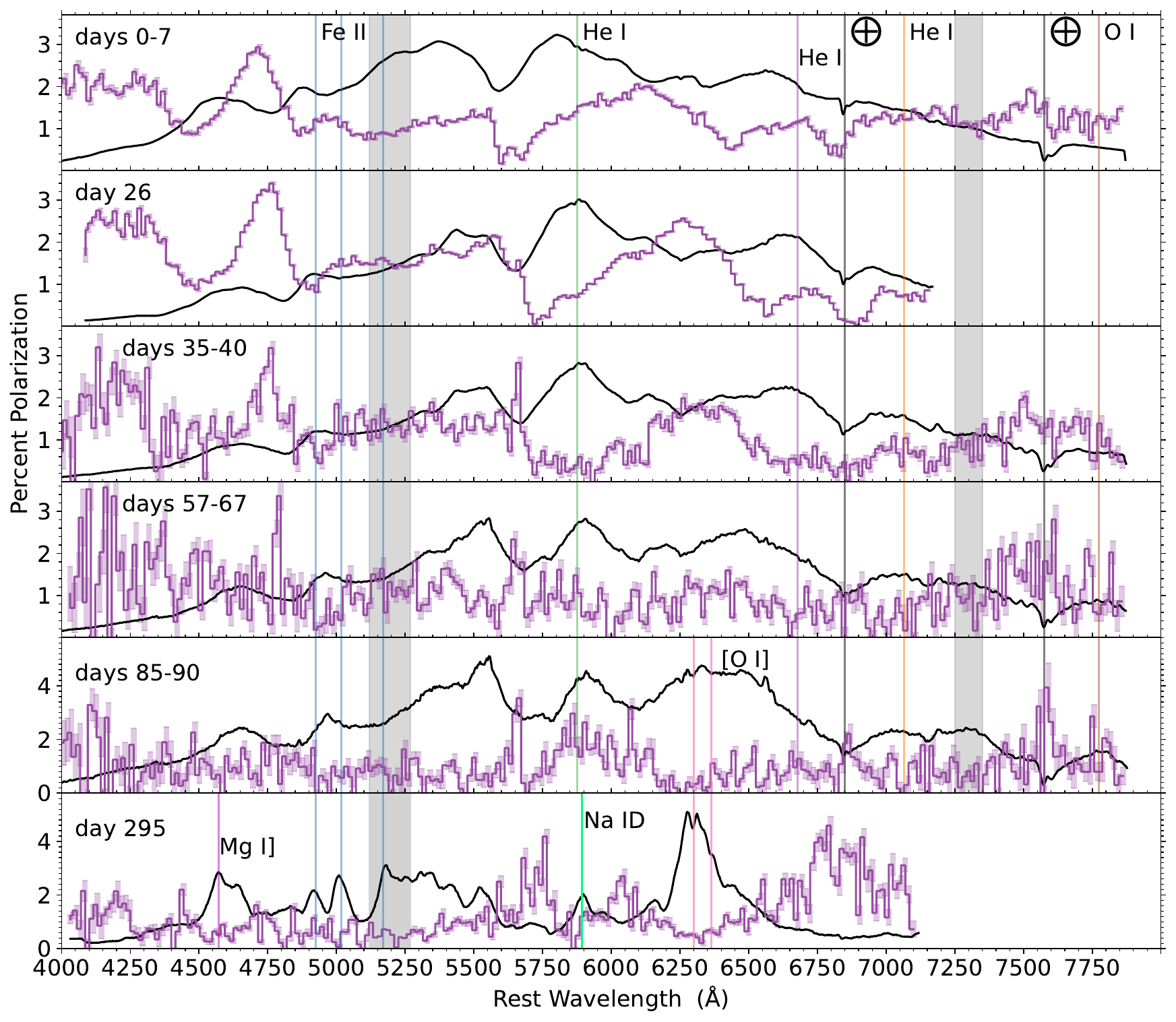}
    \caption{Flux and polarization spectra for our SN~2012au observations. The relative flux spectra are in black and polarization spectra (binned to 15 \AA) are in purple with error bars shaded. Polarization spectra are not ISP or continuum-subtracted; we note that from reddening considerations the maximum possible ISP is ~0.66\% (see \S~\ref{sec: ISP}). The rest wavelengths of prominent features in the spectra are marked by vertical lines and color-coded according to atomic species. The grey shaded regions span the ranges where we measured the continuum polarization for each epoch (\S~\ref{sec: continuum}).}
    \label{fig:specpol}
\end{figure*}

\section{ISP estimate}\label{sec: ISP}

When conducting SN~polarization analysis, we must take into consideration the potential contributions from interstellar polarization (ISP) in both the Milky Way and the host galaxy. The combined ISP can be oriented in such a way as to either add to or subtract from the object's total intrinsic polarization. 

\begin{figure}
    \centering
    \includegraphics[width=\columnwidth]{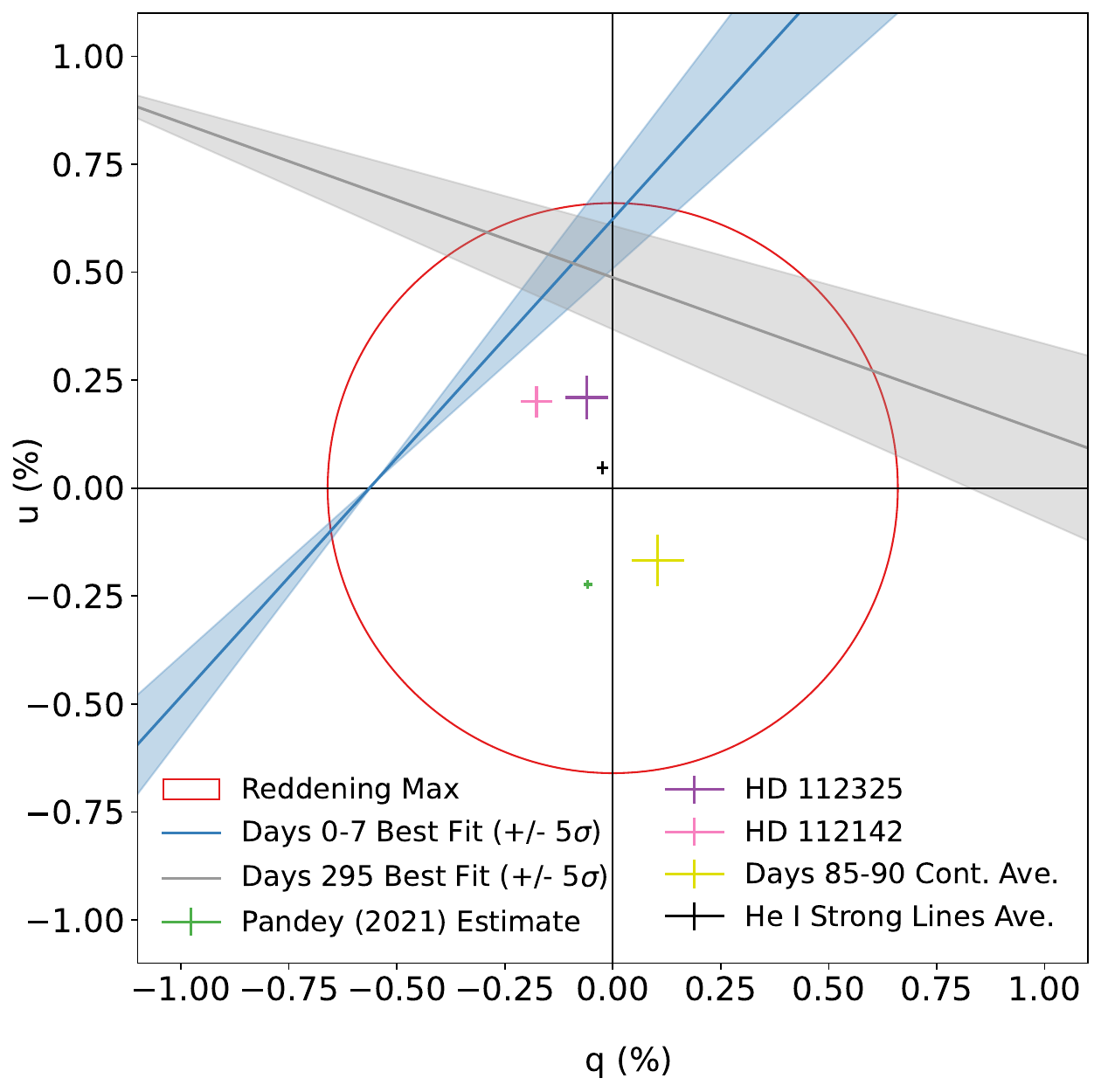}
    \caption{ISP estimates (\S~\ref{sec: ISP}) displayed on the $q-u$ plane for comparison. The ISP magnitude calculated from the Serkowski reddening maximum relationship is shown as a red circle to indicate no specific PA. The pink cross is the value of probe star HD 112142. The purple cross represents the value of the most distant probe star HD 112325. The small green cross represents the estimate presented in \citet{Pandey21}. The yellow cross is the estimate we obtain from our days $85-90$ continuum region (\S~\ref{sec: continuum}). The sizes of these crosses denote the uncertainty of each estimate. The solid blue and grey lines are the best fit lines to our data from days $0-7$ and 295, respectively, with the 5$\sigma$ errors shown as shaded regions around them. 
    }
    \label{fig:ISP_est}
\end{figure}

The correlation between ISP and reddening along the line of sight found by \cite{Serkowski75}, $p_{\textrm{max}} (\%) \leq 9 \times E(B-V)$, provides one method of estimating the maximum total ISP, including the contribution from the host galaxy. The total reddening quoted for SN 2012au by \citet[][$E(B-V) = 0.063$ mag $\pm$ 0.01 mag]{MiliD13} includes both the Milky Way value found by \citet[][$E(B-V) = 0.043$ mag]{schlafly2011} and that of the host galaxy calculated from the equivalent width of the Na \textsc{i} D absorption in their spectra ($E(B-V) = 0.02$ mag $\pm$ 0.01 mag). Using the total $E(B-V)$ value in the Serkowski relation yields $p_{\textrm{max}}\leq 0.567\% \pm 0.09\%$. Applying the maximum uncertainty to this value, we estimate the upper limit ISP magnitude to be $p_{\textrm{max}}=0.66\%$. However, this method does not provide any information on the Stokes $q$ and $u$ components of $p_{max}$, and thus the PA of this ISP estimate is unknown. Without an angle measurement for this estimate, we represent it graphically as a red circle with a radius that corresponds to this magnitude, as shown in Figure \ref{fig:ISP_est}.

However, determining the exact level of ISP can often be difficult; past studies have used numerous methods to estimate the ISP contribution and thus constrain their measurements \citep[e.g.,][]{Reilly16, Stevance20}. We also explored multiple methods for constraining the ISP; we detail these in Appendix \ref{sec: Appendix ISP} and display the resulting estimates in Figure \ref{fig:ISP_est}. Rather than choosing one single estimate and subtracting this value from all our data, we chose to conduct our polarization analysis in the $q-u$ plane and consider significant features to be those with polarization values (or changes in polarization) greater than any of these estimates. Since the Serkowski reddening ISP estimate yielded the highest polarization value ($p \leq 0.66\%$), we display this ISP estimate in subsequent $q-u$ figures as a red circle (as in Fig.~\ref{fig:ISP_est}) for comparison to the data. We do this to acknowledge the discrepancies among the PA values from all the other ISP estimates explored in Appendix \ref{sec: Appendix ISP}.  

Additionally, calculating the ISP using the Serkowski-reddening method assumes that the combined effect of the dust in both galaxies behaves like a Serkowski law as a function of wavelength \citep{Serkowski75}, which is only true in certain limited circumstances. In Appendix ~\ref{sec: Appendix Serk.} we explore this problematic assumption by analyzing a sample of 625 combinations of Milky Way and host ISP. In this work, we vary the host galaxy's peak wavelength, peak polarization value, and polarization angle, as well as the Milky Way's peak wavelength. We find that even if the dust in each galaxy individually behaves as a Serkowski law, the combined effects result in a function somewhat resembling a Serkowski curve about half the time and closely resembling a Serkowski curve only 1/3 of the time. From this we conclude that applying a Serkowski law as a wavelength-dependent ISP estimate may not yield an accurate representation of the ISP. However, adding the host and Milky Way peak polarization values (assuming they have the same PA, as we do here) results in a reasonable wavelength-independent upper limit to the ISP magnitude.

\begin{figure}
    \centering
    \includegraphics[width=\columnwidth]{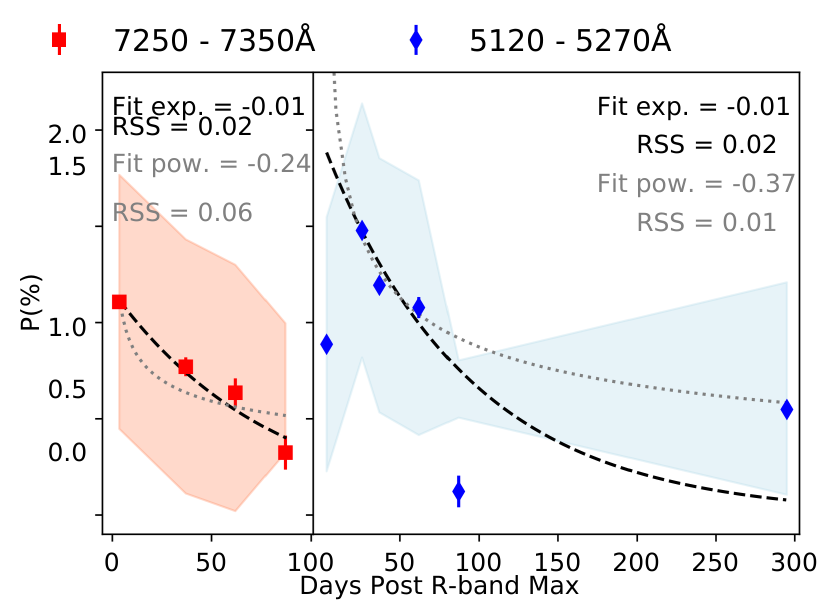}
    \caption{Continuum polarization for SN~2012au by observation day. Error-weighted means of the total polarization for the narrow continuum regions defined in \S~\ref{sec: continuum}, tabulated as $\overline{p}$ in Table~\ref{Tab:continuum_regions}, are displayed as the red squares (7250--7350~\AA) and blue diamonds (5120--5270~\AA). Corresponding exponential fits are shown as black dashed lines with the corresponding exponent and residual sum of squares (RSS) displayed in black. For comparison, we also show power law fits with grey dotted lines and the power and RSS displayed in grey. Shaded regions around the data points represent the range of polarization magnitudes under different ISP assumptions (\S~\ref{sec: continuum}; Table~\ref{Tab:continuum_regions}). Exponential fits were not applied to these ISP-adjusted points because each point is calculated using a different ISP PA, dissolving any relation the points might have to one another over time.}
    \label{fig:cont_est}
\end{figure}

\section{Continuum estimates} \label{sec: continuum}

\begin{table*}
 \centering
 \begin{tabular}{l c c c c c c c c c c c c }
    \hline
    Epoch (days) & $\overline{q}~(\%)$ & $\sigma_{q}~(\%)$ & $\overline{u}~(\%)$  & $\sigma_{u}~(\%)$ & $\theta ~(\degr)$ & $\sigma_{\theta}~(\degr)$ & $\overline{p}~(\%)$ & $\sigma_{p}~(\%)$ & $\overline{p}_\textrm{min}~(\%)$ & $\overline{p}_\textrm{max}~(\%)$ & $E_{min}$ & $E_{max}$\\
    \hline
    \multicolumn{13}{c}{Blue continuum region,  5120--5270~\AA}\\ 
    \hline
    0--7 & 0.04 & 0.01\tablenotemark{a} & 0.89 & 0.01\tablenotemark{a} & 43.6 & 0.3 & 0.89 & 0.01\tablenotemark{a}& 0.23 & 1.55 & 0.9 & 0.7\\
    26 & 0.77 & 0.01\tablenotemark{a} & 1.26 & 0.01\tablenotemark{a} & 29.2 & 0.1 & 1.48 & 0.01\tablenotemark{a} & 0.81 & 2.13  & 0.8 & 0.6 \\
    35--40 & 0.83 & 0.03\tablenotemark{*} & 0.86 & 0.03\tablenotemark{a} & 23.1 & 0.8 & 1.19 & 0.03\tablenotemark{a} & 0.53 & 1.85 & 0.9 & 0.7 \\ 
    57--67 & 0.72 & 0.06 & 0.80 & 0.05 & 24.0 & 1.5 & 1.08 & 0.05 & 0.42 & 1.74 & 0.9 & 0.7 \\ 
    85--90 & 0.15 & 0.08 & 0.01 & 0.08 & 1.4 & 22.9 & 0.12 & 0.08 & 0.51\tablenotemark{b} & 0.81 & 1.0 & 0.9 \\
    295 & 0.15 & 0.03\tablenotemark{a} & 0.42 & 0.04\tablenotemark{a} & 35.0 & 2.2 & 0.55 & 0.04\tablenotemark{a} & 0.22\tablenotemark{b} & 1.10 & 1.0 & 0.8 \\
    \hline
    \multicolumn{13}{c}{Red continuum region,  7250--7350~\AA}\\ 
    \hline
    0--7 & 0.42 & 0.02\tablenotemark{a} & 1.02 & 0.02\tablenotemark{a} & 33.8 & 0.5 & 1.11 & 0.02\tablenotemark{a} & 0.45 & 1.77 & 0.9 & 0.7 \\
    26\tablenotemark{c} & 0.20 & 0.05 & 0.87 & 0.05 & 13.7 & 1.1 & 0.89 & 0.05 & 0.23 & 1.55 & 0.9 & 0.7 \\
    35--40 & 0.40 & 0.05 & 0.66 & 0.05 & 29.3 & 1.8 & 0.77 & 0.05 & 0.11 & 1.43 & 0.9 & 0.7 \\ 
    57--67 & 0.15 & 0.07 & 0.62 & 0.07 & 38.0 & 3.2 & 0.63 & 0.07 & 0.02\tablenotemark{b} & 1.30 & 1.0 & 0.8 \\ 
    85--90 & 0.07 & 0.09 & $-$0.33 & 0.09 & 140.6 & 8.4 & 0.32 & 0.09 & 0.32\tablenotemark{b} & 1.00 & 1.0 & 0.8 \\
    \hline
 \end{tabular}
 \caption{Error-weighted mean polarization for each continuum region and observation period (denoted by days post $R$-band maximum). We calculated the minimum and maximum polarization magnitudes assuming the extreme cases of the upper limit ISP (\S~\ref{sec: ISP}). Corresponding ellipticities for the minimum and maximum polarization averages were inferred from \citet{Hoflich91}, where $E = 1$ represents a perfectly spherical ejecta model.}
 \tablenotetext{a}{Calculated value is smaller than our adopted systematic uncertainty ($0.05\%$).}
 \tablenotetext{b}{Value calculated using a direct subtraction of ISP results in a PA that is 90\degr~different from the measured value (180\degr~away in the $q-u$ plane).}
 \tablenotetext{c}{Values were calculated by other means (\S~\ref{sec: line polarization}).}
 \label{Tab:continuum_regions}
\end{table*}

In order to quantify the polarization behavior of the SN~continuum, we employed two different techniques. First, we isolated two spectral regions (shown as the grey shaded regions in Fig.~\ref{fig:specpol}) to derive estimates for the mean continuum polarization at each observation date, which we tabulate in Table \ref{Tab:continuum_regions}. Our choice of continuum regions was influenced by the analyses of  \citetalias{Maund07} and \citetalias{Reilly16} for similar SNe Ib. However, due to the complex nature of the polarization spectra of SN~2012au, we were able to use only a limited portion of the continuum regions defined in these previous works; for example, we investigated a bluer continuum region (5120--5270~\AA) than any used in \citetalias{Maund07}, \citetalias{Maund09}, or \citetalias{Reilly16}. We chose each region such that it contained little fluctuation in polarization and no corresponding feature in the flux spectrum. We then used the Stokes $Q$ and $U$ fluxes in each of these two continuum regions to calculate the error-weighted means $\overline{q}$ and $\overline{u}$, the resulting total polarization $\overline{p}$, and all associated uncertainties (Table~\ref{Tab:continuum_regions}). 

In Figure \ref{fig:cont_est}, we display the time variation of the mean total polarization for each continuum region. The minimum continuum polarization value from either region at any time is $\bar{p} = 0.12\% \pm 0.08\%$ from the blue region at days 85--90, while the maximum is $\bar{p} = 1.48\% \pm 0.05\%$ from the blue region at day 26. We tested different functions to describe the continuum behavior over time, using all red continuum points and any combination of the blue continuum points (Figure \ref{fig:cont_est}).
We found that an exponential curve described the data slightly better than a power law, with the best option for both continuum regions constructed from only the averages of days 26, 35--40 and 57--67 for the blue continuum region. We show both fits in Fig. \ref{fig:cont_est}. When calculating these fits, we did not subtract ISP
because of the uncertainty in the true ISP PA (\S~\ref{sec: ISP}). However, under the assumption that at later times the majority of the continuum polarization is due to the ISP, we averaged the red and blue continuum region values for our days 85--90 observation to calculate a single value for an ISP estimate (\S~\ref{sec: ISP}; Figure \ref{fig:ISP_est}). This ISP estimate agrees well in magnitude but not in angle with the other probe star estimates, and overall deviates the most from the general trend of the other estimates. 

\begin{figure*}
    \centering
    \includegraphics[width=.85\textwidth]{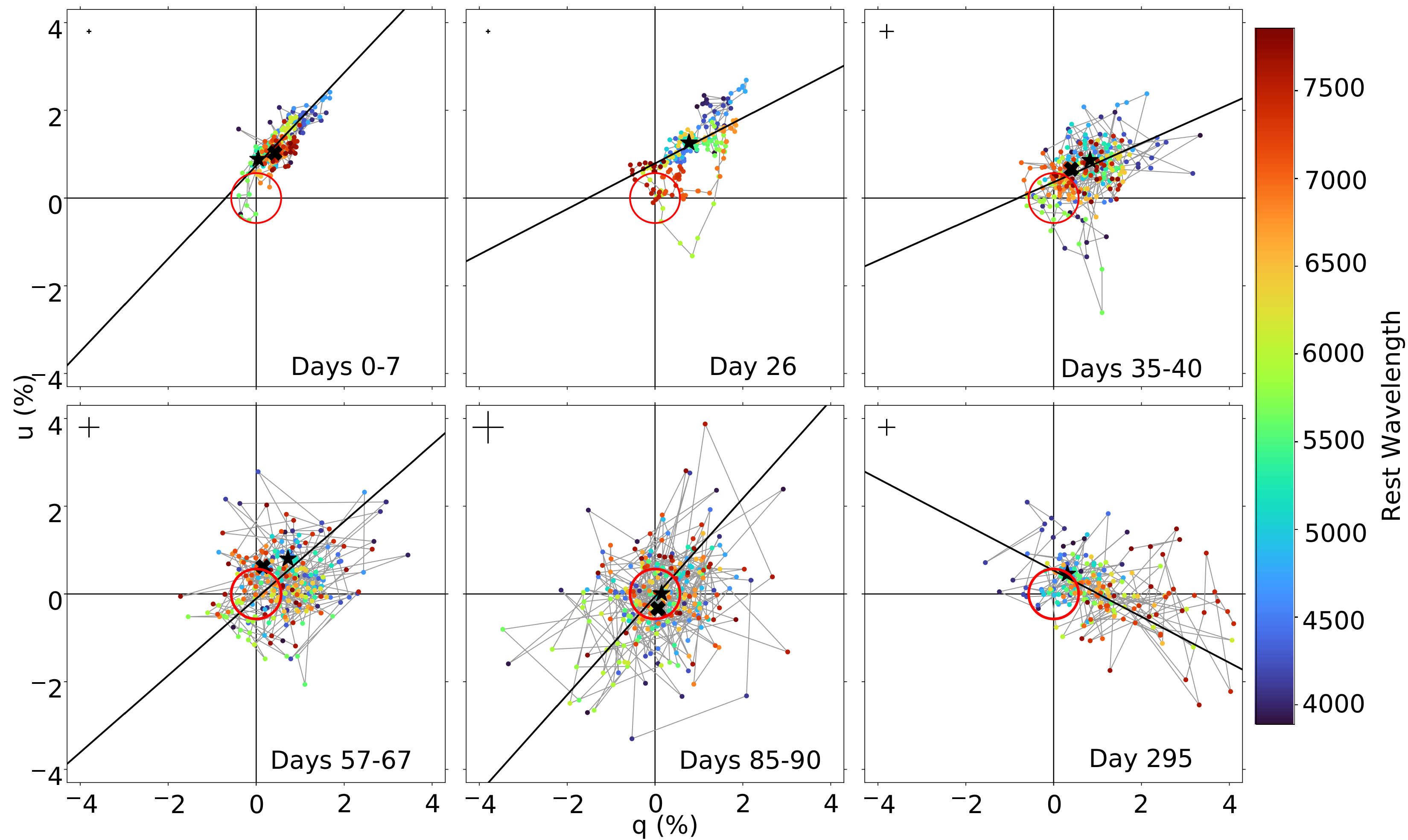}
    \caption{Full-spectrum (3890--7855 \AA, binned to 15 \AA) Stokes $q$ and $u$ data for all epochs of SN~2012au, color-coded by wavelength. Black diagonal lines represent the error-weighted best fit in each epoch to all the data excluding major line regions (He \textsc{i} $\lambda$5876, $\lambda$7065 and Fe \textsc{ii} $\lambda$$\lambda$$\lambda$4924, 5018, 5169; \S~\ref{sec: continuum}). The error-weighted average of points for the blue continuum region (5120--5270 \AA) is displayed as a black star and for the red continuum region (7250--7350 \AA) as a black X, except for day 26, where it is designated by a grey X as it was calculated by other means (\S~\ref{sec: line polarization}). Our adopted upper-limit ISP magnitude is plotted as a red circle around the origin. The average uncertainty over all points is displayed as a cross at the top left of each diagram.}
    \label{fig:all_QU}
\end{figure*}

In order to compare the intrinsic level of polarization in SN~2012au with measurements from other studies, we also tabulate  
the minimum and maximum error-weighted mean continuum polarization by adjusting our observed values to reflect the range of possibilities for the uncertain ISP contribution ($p \leq 0.66\%$; \S~\ref{sec: ISP}). We do so by applying two extreme assumptions for the ISP PA (with no assigned uncertainty on the ISP estimate itself). In case I, the ISP has the same PA as the average continuum PA for each estimate (i.e., for each continuum region and each observation epoch). In this case, we can directly \textit{subtract} the upper limit ISP value from the average continuum polarization to calculate a minimum potential intrinsic continuum polarization ($\%\overline{p}_\textrm{min}$). In case II, the ISP PA is perpendicular to the intrinsic PA (or 180\degr away in the $q-u$ plane), so we calculate the potential maximum intrinsic continuum polarization ($\%\overline{p}_\textrm{max}$) by \textit{adding} the upper limit ISP value to each observed continuum magnitude. We display the resulting $\%\overline{p}_\textrm{min}$ and $\%\overline{p}_\textrm{max}$ values 
for both continuum regions and each observation in Table \ref{Tab:continuum_regions}, and we display the range between the two as the shaded regions in Figure \ref{fig:cont_est}. It is worth noting that in Figure \ref{fig:cont_est} the blue region continuum estimate for days 85--90 is outside the shaded region, due to the positive nature of the (debiased) total polarization calculation when both the $q$ and $u$ components are negative with small uncertainties. Physically, not all of these ISP PA scenarios can be true at the same time, but these possibilities allow us to place reasonable limits on the intrinsic continuum polarization. From this analysis we find that the intrinsic continuum polarization for both regions, across all observations, could potentially range from $0.02\%-2.13\%$. Furthermore, we compared the $R$-band imaging polarimetry measurements made by \citet{Pandey21} at day 27 ($p = 1.32\%$) and day 47 ($p = 0.13\%$) to our red region measurements for the corresponding days (26 and 35--40, respectively) and found they both fall within our estimated minimum and maximum values (Table~ \ref{Tab:continuum_regions}). 

It is common practice in polarimetric studies to compare continuum polarization magnitudes to the electron-scattering models of \citet{Hoflich91} to deduce the level of asphericity of the ejecta. We made this comparison for the range of continuum values taking the extreme ISP cases into account ($p = 0-2.05\%$; the uncorrected polarization magnitudes fall within this range, with $p = 0.12\%-1.48\%$) and list the corresponding ellipticities in Table~\ref{Tab:continuum_regions}. From these model comparisons, we deduce that on average, the ejecta of SN~2012au may deviate from spherical symmetry at the level of $12\%$. Additionally, the general decreasing trend over time of SN~2012au's continuum polarization (regardless of ISP PA; Fig.~\ref{fig:cont_est}) is more commonly described by a prolate rather than an oblate ellipsoid in the \citet{Hoflich91} models.  

We note that the majority of average continuum polarization magnitudes (with and without ISP correction) for SN~2012au are greater than those observed in SN~1987A (the original comparison for the \citealt{Hoflich91} models). They are also
greater than the continuum estimates reported for the SNe Ib studied by \citetalias{Maund07}, \citetalias{Maund09}, and \citetalias{Reilly16}. We will discuss the implications of these comparisons in Section \ref{sec: discussion}. 

However, it is likely that the assumptions made by \citet{Hoflich91} were an oversimplification of the true nature of these SN photospheres. \citet{Dessart24} modeled the continuum polarization behavior of SN Type II-P in greater detail 
 and found that anisotropies in both the inner and fast-moving parts of the ejecta, as well as the explosion energy itself, are key factors producing higher polarization levels. This study demonstrated some scenarios in which the continuum polarization rises after the explosion until nearing the end of the photospheric phase, when it slightly decreases before jumping to a peak value at the transition from photospheric to transitional-nebular phase. It then decreases throughout the transitional-nebular phase, often at a rate of $1/t^{2}$ due to geometrical dilution. We explore this potential continuum trend for the transitional-nebular phase of SN~2012au in Section~\ref{subsec: epoch6 polarization}.
 
Two of the models presented by \citet{Dessart24} (e2ni1b1/e1ni1 and e2ni1b2/e1ni1, which only differ by the mass of the outer Ni shell) closely match the trend of the blue continuum averages we found in SN~2012au (Figure~\ref{fig:cont_est}). Unfortunately, we did not observe the SN~between 100--295 days, which would have provided a better model comparison. We also note that these continuum trends may have different implications for SN~2012au than for the SNe Type II-P that form the basis for these models. Further investigation of the implications of these models for other SESNe is a tantalizing subject but beyond the scope of this paper. 

The second continuum analysis technique we employed was to calculate global continuum fits (GCF) to the data in the $q-u$ plane to derive a better estimate for the general continuum position angle (PA) and dominant axis of each of our spectra \citep[Fig. \ref{fig:all_QU};][]{WW08}. We obtained these via an error-weighted fit to all our data at each epoch, excluding obvious line regions based on inspection of the spectra. We always excluded the wavelength regions that exhibited obvious line polarization surrounding the rest wavelengths of the helium and iron line features labeled in Figure \ref{fig:specpol}. 

We used this broad definition of the continuum under the assumption that the continuum polarization is the strongest underlying influence to the overall PA. Therefore, using as much data as possible gives us a better estimate of the GCF slope in the $q-u$ plane, from which we derive the average PA across the spectrum (GCF$\theta$). The $\chi^{2}$ and $\theta$ values we found for each epoch's GCF are recorded in Table~\ref{Tab:continuum_q-u_bfit}. As a check, we also plotted the averages of the smaller blue and red continuum regions (discussed above) in $q-u$ space and found that in most cases these coincide well with the GCF lines (except for the red continuum region estimates for days 57--67 and days 85--90, for which the data are less linear and the GCF has larger $\chi^2$ values (Fig.~\ref{fig:all_QU}; Table~\ref{Tab:continuum_q-u_bfit}). Going forward in our analysis, we adopt the calculated GCF$\theta$ as the dominant axis for each epoch of data. This allows us to directly compare differences in angle among polarization features within an epoch, and also analyze changes over time without subtracting an ISP estimate. Additionally, we use each GCF to identify the spectropolarimetric classification (SP type) for each observation according to the system introduced by \citet{WW08}, which we show in the final column of Table \ref{Tab:continuum_q-u_bfit}. We discuss these classifications further in the next section.
 
\begin{table}
 \centering
 \begin{tabular}{l c c c c c c}
    \hline
    Epoch & Slope & $\sigma_{\textrm{Slope}}$ & $\chi^2$ & GCF$\theta$ & $\sigma_{\theta}$ & SP \\
    (days) &  & & & $(\degr)$ & $(\degr)$ & Type\\
    \hline
    0--7 & 1.05 & 0.07 & 0.03 & 23.3 (203.3) &  1.0 & D0 \\
    26 & 0.51 & 0.04 & 0.02  & 13.7 (193.7) & 1.1 & L \\
    35--40 & 0.44 & 0.07 & 0.13 & 12.0 (192) & 1.8 & D1 \\ 
    57--67 & 0.87 & 0.11 & 0.37 & 20.6 (200.6) & 1.9 & N1 \\ 
    85--90 & 1.11 & 0.13 & 0.57 & 24.1 (204.1) & 1.7 & N1 \\
    295 & -0.40 & 0.07 & 0.15 & 168.9 & 2.3 & D1 \\ 
    \hline  
 \end{tabular}
 \caption{Global continuum fit (GCF) parameters (\S~\ref{sec: continuum}) and inferred SP types (\S~\ref{sec: line polarization}) for our observations of SN~2012au. Epochs are specified by days post $R$-band peak brightness. In addition to the calculated GCF$\theta$ values, we list} adjusted angles (+ 180$\degr$) corresponding to the values shown in Figures \ref{fig:e1_polar_plot} and \ref{fig:polar_plots}.
  \label{Tab:continuum_q-u_bfit}
\end{table}

\section{Line Polarization}\label{sec: line polarization}

\begin{table*}
 \centering
 \begin{tabular*}{0.75\textwidth}{c c c c c c c c c c }
    \hline
    Epoch & ${p}$ & $\sigma_{p}$ & $\theta$ & $\sigma_{\theta}$  & ${\lambda}_{p}$ & $v_{p}$ & $v_{flux}$\tablenotemark{d} & $\sigma_v$ & SP Type\\
    (days) & $~(\%)$ & $~(\%)$ & (\degr) & (\degr) & (\AA) & (km s$^{-1}$) & (km s$^{-1}$) & (km s$^{-1}$) & \\
    \hline
    \multicolumn{10}{c}{He \textsc{i} $\lambda$5876 (Green) }\\ 
    \hline
    0--7 & 1.4 & 0.03\tablenotemark{a} & 130.8 & 0.6 & 5650 & $-$11530 & $-$14590 & 420 & D1/L \\
    26 & 2.58 & 0.03\tablenotemark{a} & 135.8 & 0.2 & 5670 & $-$10510 & $-$11530 & 390 & L \\
    35 -- 40 & 3.48 & 0.14 & 137.3 & 1.4 & 5670 & $-$10510 & $-$10510 & 370 & L \\
    57--67 & 2.88 & 0.18 & 138.8 & 2.6 & 5650 & $-$11530 & $-$10000 & 370 & L \\
    85--90 & 3.69 & 0.32 & 96.3 & 3.1 & 5670 & $-$10510 & $-$10510 & 370 & L \\
    85--90\tablenotemark{c} & 3.05 & 0.39 & 120.0 & 3.1 & 5870 & $-$310 & $-$4900 & 360 & 
    N1 \\
    \hline
    \multicolumn{10}{c}{He \textsc{i} $\lambda$6678 (Purple) }\\ 
    \hline
    0--7 & 0.63 & 0.04\tablenotemark{a} & 115.7 & 1.9 & 6430 & $-$11130 & $-$14280 & 320 & L\tablenotemark{b}\\
    26 & 1.02 & 0.02 & 150.2 & 0.5 & 6550 & $-$5750 & $-$10680 & 340 & D1\\
    35--40 & 1.22 & 0.08 & 148.7 & 1.9 & 6540 & $-$6200 & $-$10240 & 320 & D1\\
    \hline
    \multicolumn{10}{c}{He \textsc{i} $\lambda$7065 (Orange) }\\ 
    \hline
    0--7 & 0.78 & 0.04\tablenotemark{a} & 130.2 & 1.5 & 6810 & $-$10820 & -- & 310 & L\tablenotemark{b}\\
    26 & 0.98 & 0.02\tablenotemark{a} & 130.0 & 0.7 & 6870 & $-$8270 & -- & 330 & L \\
    35--40 & 1.14 & 0.09 & 104.6 & 2.6 & 6960 & $-$4460 & -- & 300 & L \\
    57--67 & 1.50 & 0.18 & 114.5 & 3.6 & 6820 & $-$10400 & -- & 310 & N0 \\
    85--90 & 1.88 & 0.35 & 147.7 & 5.2 & 6870 & $-$8270 & -- & 350 & N1 \\
    \hline
    \multicolumn{10}{c}{Fe \textsc{ii} $\lambda\lambda\lambda$4924, 5018, 5169 (Blue, $\lambda_{rest} = 5018$)}\\ 
    \hline
    0--7 & 2.23 & 0.05 & 201.6 & 0.7 & 4720 & $-$17630 & $-$14820 & 510 & L \\
    26 & 1.93 & 0.03\tablenotemark{a} & 203.8 & 0.5 & 4770 & $-$14650 & $-$12430 & 470 & D1 \\
    35--40 & 1.99 & 0.16 & 204.7 & 2.4 & 4770 & $-$14650 & $-$11830 & 470 & L \\
    57--67 & 2.29 & 0.34 & 200.6 & 4.3 & 4800 & $-$12850 & $-$10630 & 440 & L\\
    \hline
    \multicolumn{10}{c}{[O \textsc{i}] $\lambda\lambda$6300, 6330 (Pink, $\lambda_{rest} = 6300$)}\\ 
    \hline
    0--7 & 0.95 & 0.03 & 209.7 & 0.9 & 6100 & $-$9520 & $-$6660 & 340 & N1 \\
    57--67 & 1.77 & 0.18 & 131.9 & 3.4 & 6090 & $-$9990 & $-$4760 & 340 & N1 \\
    85--90 & 2.92 & 0.38 & 113.6 & 3.8 & 6080 & $-$10470 & $-$950 & 340 & N1 \\
    \hline
    \multicolumn{10}{c}{O \textsc{i} $\lambda$7774 (Brown) }\\ 
    \hline
    0--7 & 0.85 & 0.05 & 204.3 & 2.0 & 7530 & $-$9410 & -- & 320 & N0/L\\
    35--40 & 1.31 & 0.15 & 194.8 & 3.6 & 7510 & $-$10180 & -- & 320 & L \\
    57--67 & 2.49 & 0.33 & 184.6 & 5.0 & 7620 & $-$5940 & -- & 310 & N1 \\
    85--90 & 4.25 & 0.90 & 217.8 & 4.4 & 7590 & $-$7100 & -- & 280 & L \\
    \hline
 \end{tabular*}
 \caption{Photospheric phase line polarization details for SN~2012au by epoch. Spectral lines are identified by rest wavelength as well as a color assigned to that line polarization data in Figures \ref{fig:specpol} and \ref{fig:polar_plots}. For each spectral line we list the the polarization properties and corresponding wavelength and velocity at the peak of the line. Then we note the velocity of the corresponding absorption/emission feature in the flux spectra ($v_{flux}$), and the corresponding SP type \citep{WW08} based on the line's behavior in $q-u$ space.} Polarization and PA ($\theta$) values shown are the resulting quantities after continuum subtraction (\S~\ref{sec: line polarization}). SP types are assigned with respect to that epoch's GCF (\S~\ref{sec: continuum}).
 \tablenotetext{a}{Observed value is smaller than our adopted systematic uncertainty ($0.05\%$).}
 \tablenotetext{b}{Line data do not extend past the ISP upper limit in $q-u$ space after continuum subtraction, suggesting SP type N0 (no intrinsic line polarization).}
 \tablenotetext{c}{Secondary polarization peak for line region.}
 \tablenotetext{d}{Values not reported when line feature is indistinguishable in the flux spectrum.}
 \label{Tab:phot_line_pol}
\end{table*}

To isolate the polarization behavior of the line regions in SN~2012au, we removed our estimates of continuum polarization (\S~\ref{sec: continuum}) from the data in these regions via vector subtraction. From this point on, any quoted line polarization values for photospheric phase features were calculated relative to the average polarization of the nearest continuum region for the corresponding observation date (Table \ref{Tab:continuum_regions}). For the day 26 observation, for which a red continuum region average was not available (Fig.~\ref{fig:specpol}), we adopted a continuum reference point on the GCF line for this observation with a polarization magnitude interpolated from the exponential fit to the red continuum data (Fig.~\ref{fig:cont_est}). We did not subtract any continuum from the line polarization values quoted for the day 295 observation, due to the uncertainty in the behavior of the continuum polarization at such a late time in recent model predictions \citep{Dessart24}. We also did not explicitly subtract the upper limit ISP from any data (due to our uncertainty in its PA; \S~\ref{sec: ISP}), under the premise that the continuum level we remove has the true ISP value embedded in it, and that the observed polarization signals are due only to effects of occulting material blocking portions of the ejecta that are polarized by electron scattering. This is the approach taken by other SN~Ib studies, e.g., \citetalias{Tanaka09} and \citetalias{Tanaka12}.

Our photospheric phase spectra (Fig.~\ref{fig:specpol}) show distinct polarization features corresponding to He \textsc{i}, Fe \textsc{ii} and O \textsc{i} absorption (and/or emission) flux signatures (\S~\ref{sec: spectral_evolution}). The polarization magnitude for these and additional line features seen in SN~2012au all have magnitudes $p > 0.85\%$ (after continuum subtraction) throughout most of the observed epochs; we list these and other important line traits in Table~\ref{Tab:phot_line_pol}. These line features exhibit varying behaviors with respect to the dominant axis (\S~\ref{sec: continuum}) and orthogonal axis (perpendicular to the dominant axis) in the $q-u$ plane (Fig. \ref{fig:all_QU}), which correspond to different SP types\footnote{SP types are defined by the behavior of the data in the $q-u$ plane: N0 = data consistent with noise; N1 = data signal is greater than noise but no dominant axis can be defined; D0 = a dominant axis can be fit to majority of data by a straight line with minor axis distribution consistent with noise; D1 = a dominant axis can be identified but a straight line does not return a satisfactory fit due to data having significant distribution orthogonal to the line; L = data show significant variation in polarization magnitude and PA across a line such that a loop is created.} as defined by \citet{WW08}. In this system, our data for SN~2012au can be classified as follows. In days 0--7 the bulk of the data trend along the dominant axis, consistent with SP type D0. At day 26, our data show a linear trend in the continuum and clear $q-u$ loops across the He \textsc{i} $\lambda$5876, $\lambda$6678 and $\lambda$7065 lines, resulting in SP type L. At days 35--40 and day 295, the data cluster around a dominant axis, but with large deviations; SP type D1 is most consistent with this behavior, but we note that the angle of the dominant axis dramatically rotates ($\Delta\theta\approx 23\degr$; Table~\ref{Tab:continuum_q-u_bfit}) between these two epochs. Finally, days 57--67 and 85--90 show good signal but little evidence for a preferred axis, yielding SP type N1. We tabulate these classifications in Table~\ref{Tab:continuum_q-u_bfit}.

\begin{figure}
    \centering
    \includegraphics[width=\columnwidth]{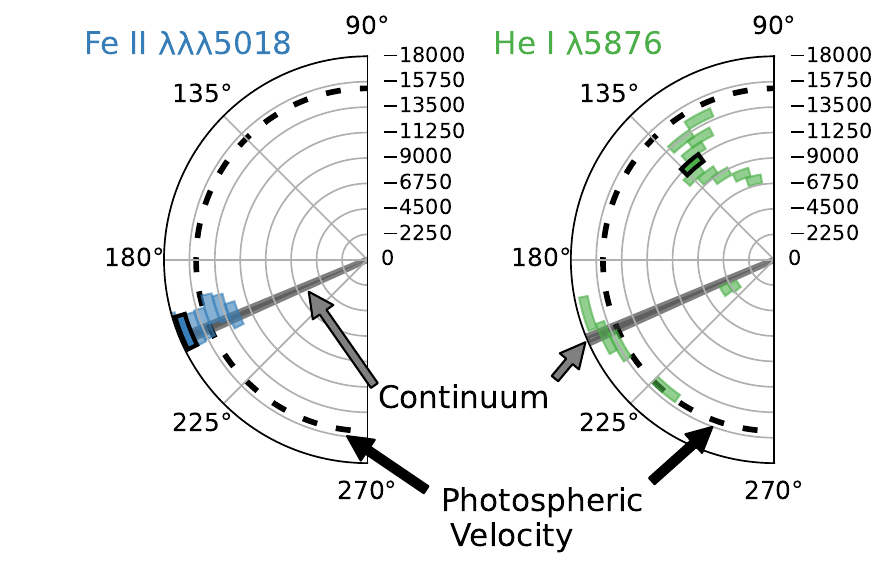}
    \caption{Polar plots showing the Fe \textsc{ii} (\textit{left}) and He \textsc{i} (\textit{right}) line regions after continuum subtraction for days 0--7. For each line region the PA ($+180\degr$ for clarity of angle variation) is plotted as a function of blueshifted velocity (on the radial axes). Data are binned to 15 \AA~ and represented by individual blocks; the angular width of each block is either $10\degr$ or the uncertainty in PA for that data point (whichever is larger for legibility). The continuum is represented by a grey wedge at the angle of the GCF slope ($+ 180\degr$) spanning the width of the calculated PA uncertainty and stretched over the whole velocity space. The black, dashed semicircle represents the photospheric velocity, measured by the shift of the Fe \textsc{ii} flux absorption minimum from rest (\S~\ref{subsec: iron flx}).}
    \label{fig:e1_polar_plot}
\end{figure}

\begin{figure*}
    \centering
    \includegraphics[width =.85\textwidth]{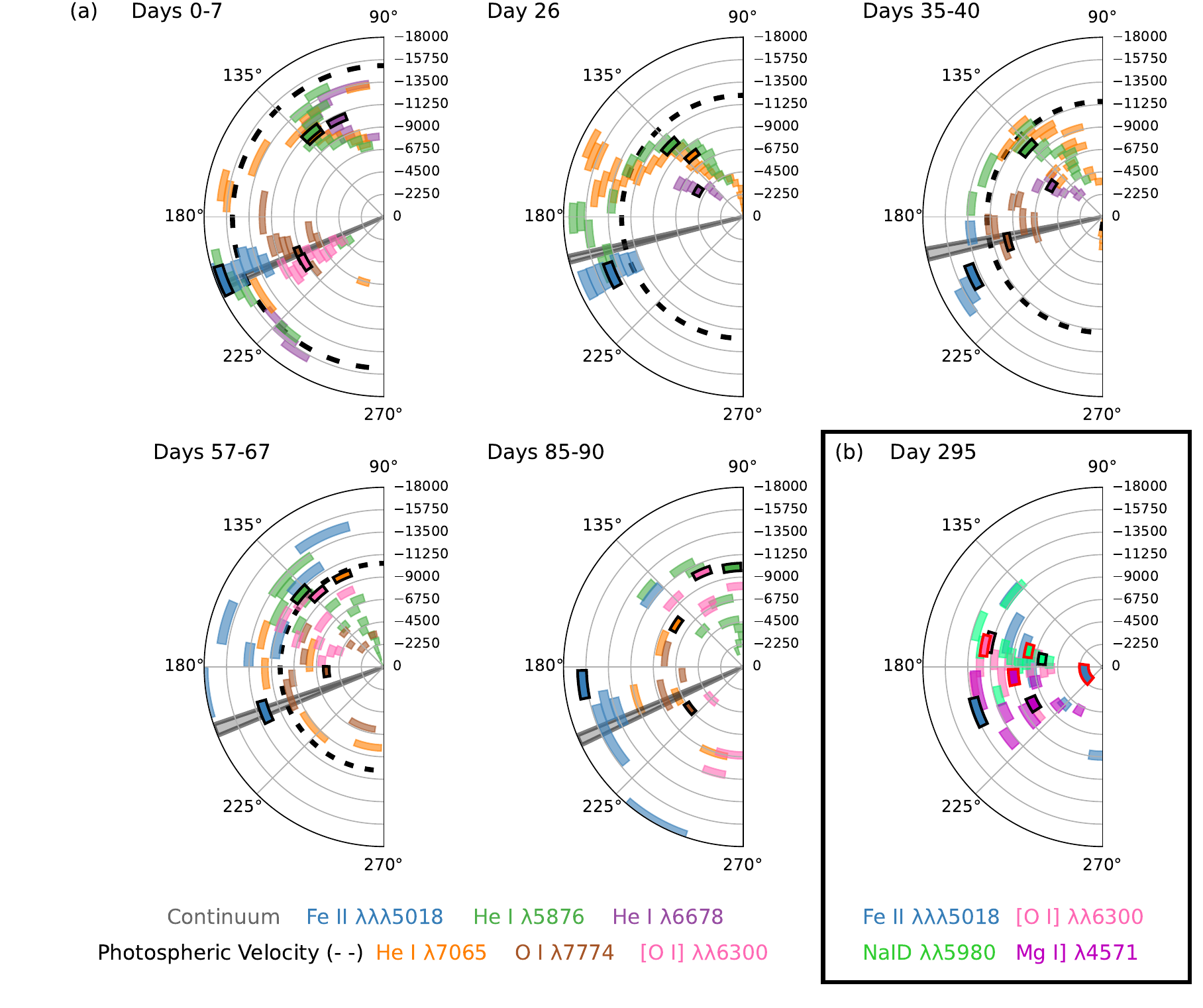}
    \caption{Polar plots of line polarization for SN~2012au throughout all (a) photospheric (days 0--90) and (b) transitional-nebular phase (day 295) epochs. All polarization values have had continuum subtracted, except for those shown on day 295. For clarity, we plot PA $+180\degr$ for different line regions as a function of blueshifted velocity (on the radial axes). The spread of the angle for each colored region is either $10\degr$ or the uncertainty in PA for that region, whichever is larger for legibility. The angle corresponding to the GCF slope (GCF$\theta$; \S~\ref{sec: continuum}) has also been adjusted by $+180\degr$ and is shown as a translucent grey wedge, spanning the width of the calculated PA uncertainty and stretched over the whole velocity space. The black, dashed semicircle represents the photospheric velocity (\S~\ref{subsec: iron flx}). Line polarization in regions for photospheric epochs (defined in  \S~\ref{sec: line polarization}) are shown for Fe \textsc{ii} multiplet 42, He \textsc{i} $\lambda$5876, $\lambda$6678, $\lambda$7065, O \textsc{i} $\lambda$7774, and [O \textsc{i}] $\lambda\lambda$ 6300, 6330, as blocks with heights determined by a 15~\AA~bin size converted to velocity for the appropriate rest wavelength. Blocks corresponding to the peak polarization are solid filled and outlined in black. The peak polarization values for He \textsc{i} $\lambda$5876 and $\lambda$7065 at days 0--7 share the same PA and thus overlap in the first panel; we show only the He \textsc{i} $\lambda$5876 block. The plot for day 295 (b) shows a different subset of line polarization, including Mg \textsc{i} $\lambda$4571 and Na \textsc{i} $\lambda\lambda$5980, 5986 for the wing regions described in \S~\ref{subsec: epoch6 polarization}, with the peak polarization of the red wing regions plotted as a single block with a red outline by taking the negative equivalent of the redward velocity shift where it appears. There is no continuum or photospheric velocity defined for this epoch.} 
    \label{fig:polar_plots}
\end{figure*}

To illustrate the polarization behavior of the line regions 
over time, we compare them in polar plots (PA as a function of velocity; Figures \ref{fig:e1_polar_plot} and \ref{fig:polar_plots}) and in $q-u$ space (Figure \ref{fig:He_lines_QU_vel}). 
Figure \ref{fig:e1_polar_plot} is a simplified polar plot of Fe \textsc{ii} multiplet 42 and He \textsc{i} $\lambda$5876 during days 0--7, demonstrating the differences in their PA behavior and the velocities of their peak line polarization. Figure \ref{fig:polar_plots} compares the PA for multiple lines at each epoch, revealing a common PA for multiple He \textsc{i} lines that is different from those of other spectral lines and the GCF$\theta$. Additionally, the PAs corresponding to the peak polarization values of the He \textsc{i} lines are significantly rotated with respect to the faster-moving material in the same lines. This PA rotation results in loops and/or origin crossings in the $q-u$ plane (Figure~\ref{fig:He_lines_QU_vel}), motivating our designation of the day 26 spectrum as SP type L. 

The Fe \textsc{ii} region has a PA that remains roughly constant over time, and the corresponding polarization signature peaks at consistently higher velocities than the peaks of the other lines (Figure~\ref{fig:polar_plots}). The polarization signatures in oxygen line regions (O \textsc{i} $\lambda$7774,  [O \textsc{i}] $\lambda\lambda$ 6300, 6330) share similar PAs to the Fe \textsc{ii} region at days 0--7, 
but these
behave differently over time. Although identifying distinct polarization signatures for each line region becomes challenging for days 57--67 and 85--90, it is clear that at these epochs there is no longer a preferred location for the bulk of the data (continuum and non-helium lines). 
However, in our transitional-nebular phase epoch (day 295), a new dominant axis arises that aligns with most of the line regions (Fig.~\ref{fig:He_lines_QU_vel}; Table~\ref{Tab:continuum_q-u_bfit}).   

We discuss the time-dependent spectropolarimetric behavior of prominent line regions in more detail below.
We combine the analysis of all line polarization features from our transitional-nebular phase (day 295) observation into a separate subsection, \S~\ref{subsec: epoch6 polarization}.

\subsection{Helium}\label{subsec: Helium}
SNe Ib typically show prominent helium signatures in their polarized spectra (\citetalias{Maund07, Tanaka09, Maund09, Tanaka12, Reilly16}), and SN~2012au is no exception (Fig.~\ref{fig:specpol}). Each of the five photospheric phase epochs shows a well defined polarization peak bluewards of He \textsc{i} $\lambda$5876, closely aligning with a prominent flux absorption feature associated with that line (\S~\ref{subsec:Helium_flux_spec}). Polarization peaks blueshifted from He \textsc{i} $\lambda$6678 and He \textsc{i} $\lambda$7065 by a similar velocity to He \textsc{i} $\lambda$5876 are clearly present at days 0--7, but less identifiable in the other photospheric epochs. Our data also show a distinct peak in polarization at $\approx4700$~\AA~for days 0--7 through days 35--40, which corresponds to a velocity shift from He \textsc{i} $\lambda$5016 similar to those of the other prominent He \textsc{i} lines. However, this polarization feature most likely has significant contributions from the overlapping Fe \textsc{ii} triplet ($\lambda$$\lambda$$\lambda$4924, 5018, 5169; \S~\ref{subsec:Iron}).

\begin{figure*}
    \centering
    \includegraphics[width=.8\textwidth]{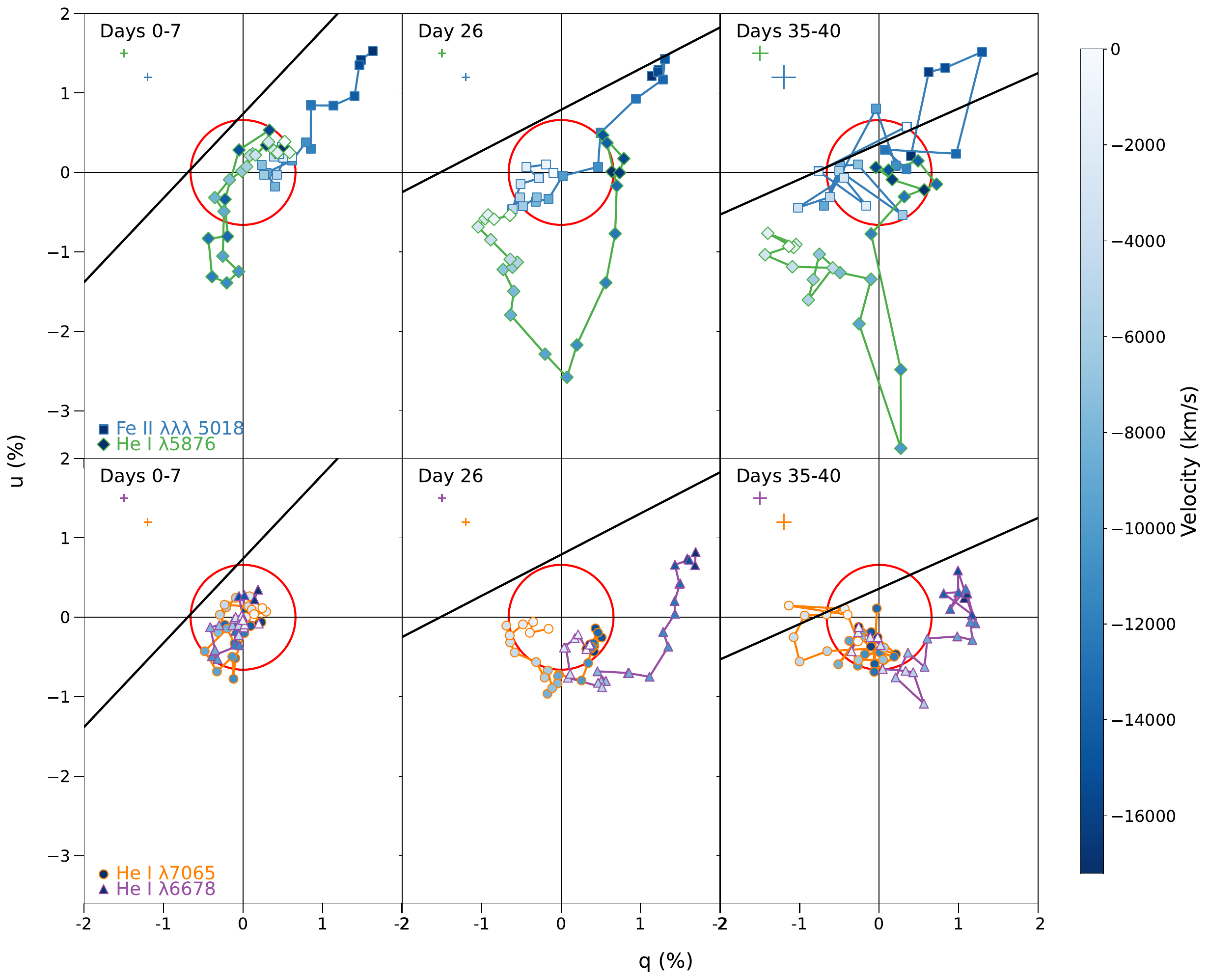}
    \caption{Early photospheric phase behavior in $q-u$ space of the He \textsc{i} $\lambda$5876 (diamonds and green lines), potentially blended He \textsc{i}/Fe \textsc{ii} $\lambda$5018 (squares and dark blue lines), He \textsc{i} $\lambda$6678 (triangles and purple lines), and He \textsc{i} $\lambda$7065 (circles and orange lines) absorption features in the velocity range of $0-18,000$ km s$^{-1}$ after continuum subtraction. Velocities are indicated by the point colors, referring to the color bar at right. The upper limit ISP is plotted as a red circle on each panel and average uncertainties for each line region are shown in the upper left of each panel. For calculated average uncertainties less than 0.05\%, we display the upper limit instrumental uncertainty of 0.05\% (\S~\ref{sec:Obs&reduction}). The GCFs at each epoch are plotted in black for comparison (\S~\ref{sec: continuum}); because of the continuum subtraction, these lines are now offset from the data points.}
    \label{fig:He_lines_QU_vel}
\end{figure*}

All our observations from day 26 through days 85--90 exhibit polarization levels of $p > 0.60\%$ (after continuum subtraction) across regions surrounding these He \textsc{i} lines. The broadening of these line regions away from well-defined peaks may be due to time-varying density and temperature profiles that influence the emergence of other neighboring spectral lines with potentially competing polarization signatures. Keeping this in mind, we chose to compare the helium lines' behavior by isolating regions around the He \textsc{i} $\lambda$6678 and He \textsc{i} $\lambda$7065 lines spanning the same velocity range as the easily identified He \textsc{i} $\lambda$5876 polarization signature (Fig. \ref{fig:He_lines_QU_vel}). 

At days 0--7 the He \textsc{i} $\lambda$5876 polarization signature has a peak magnitude of $p = 1.40 \% \pm 0.05\%$ at $v = -11,530  \pm 420$ km s$^{-1}$. The peak polarization for this line increases in magnitude and decreases in velocity throughout the photospheric phase (apart from a slight decrease in polarization at days 57--67 compared to days 35--40); it reaches a maximum polarization of $p = 3.69 \% \pm 0.32\%$ by days 85--90, located at $v = -10,510 \pm 370$ km s$^{-1}$. The peak polarization values we associate with He \textsc{i} $\lambda$6678 and He \textsc{i} $\lambda$7065 also increase over time, though the overall magnitude of each is consistently less than that of the He \textsc{i} $\lambda$5876 peak by $\approx1-2\%$. At days 57--67 and 85--90, the peak in polarization corresponding to He \textsc{i} $\lambda$5876 aligns with the blue side of the now-broadened flux absorption feature (\S~\ref{sec: spectral_evolution}). At the same time, the polarization signatures for He \textsc{i} $\lambda$7065 and potential He \textsc{i} $\lambda$6678 become harder to identify as distinct features. In general, the data for these later photospheric phase epochs are noisier and more difficult to interpret; we therefore focus here on the first three epochs and compare their behavior with that of the transitional-nebular phase epoch (day 295) in \S~\ref{sec: discussion}.

Figure \ref{fig:He_lines_QU_vel} compares all three He \textsc{i} lines (along with the potentially blended He \textsc{i} 5016/Fe \textsc{ii} 5018 feature discussed in \S~\ref{subsec:Iron}) in $q-u$ space throughout the early photospheric phase. In our first three epochs, the most prominent He \textsc{i} lines ($\lambda$6678, $\lambda$7065, and $\lambda$5876) behave similarly in $q-u$ space, undergoing significant PA rotations across their lines. By contrast, the He \textsc{i}/Fe \textsc{ii} line region mostly resides in the first quadrant and maintains a more consistent PA. At days 0--7 the line regions for $\lambda$6678, $\lambda$7065, and $\lambda$5876 all roughly have the same slope as the dominant axis defined by the GCF 
(Fig.~\ref{fig:He_lines_QU_vel}), but have a different PA than that shared by the GCF and He \textsc{i}/Fe \textsc{ii} at this time (Figs.~\ref{fig:e1_polar_plot} and \ref{fig:polar_plots}). At day 26, these lines further deviate from the GCF, all creating loop-like patterns in the $q-u$ plane. At this same epoch the He \textsc{i}/Fe \textsc{ii} region again behaves differently from the other He lines, in that it lacks a loop signature
(Fig. \ref{fig:polar_plots}). At days 35--40 the $\lambda$5876 loop has collapsed to become more linear, while the $\lambda$7065 loop is more distinctly separated from the ISP upper limit, and all three line regions still deviate from the dominant axis. Meanwhile the He \textsc{i}/Fe \textsc{ii} region develops a loop covering different PAs than those previously traced by the loops in the other lines. Under these considerations, we associate the He \textsc{i}/Fe \textsc{ii} line region more closely with the Fe \textsc{ii} triplet than with the He \textsc{i} $\lambda$5016 line, and discuss it further in the next section. 



Over the first three observations, the amount of PA rotation across the $\lambda$5876, $\lambda$6678 and $\lambda$7065 line regions decreases (Fig. \ref{fig:polar_plots}). At days 0--7 this PA changes abruptly from higher velocity to lower velocity, while at days 26 and 35--40 the rotation is smooth. For example, in Figure \ref{fig:e1_polar_plot} (days 0--7), the green rectangles representing $\lambda$5876 start out at $\approx195\degr$ at high velocities then, jump to $\approx120\degr$ at lower velocities. Figure \ref{fig:polar_plots} shows that this trend continues at days 35--40, but with a slightly smaller PA rotation. At first all three He \textsc{i} PA line profiles show the same rotation of $\approx75\degr$. At day 26 all three He \textsc{i} lines share a similar PA trend near peak polarization (Figure \ref{fig:polar_plots}). By days 35--40 all the He \textsc{i} lines span narrower velocity regions with small PA rotations, but the $\lambda$6678 PA remains offset from the other two. Although the $\lambda$6678 line creates a similar shape in $q-u$ space to $\lambda$5876 and $\lambda$7065 (as described above), this offset in PA is reflected in the fact that its path lies in a different quadrant than those of the other two lines (Fig.~\ref{fig:He_lines_QU_vel}). These similarities in the polarization behavior of He \textsc{i} $\lambda$5876 and $\lambda$7065 may indicate that other excited material, such as hydrogen or silicon, contributes to the polarization in the $\lambda$6678 region, 
or that the $\lambda$6678 polarization may arise from a different region in the ejecta than the shared scattering location of the $\lambda$5876 and $\lambda$7065 material.  
       
\subsection{Iron}\label{subsec:Iron}
We find a consistent polarization peak at ~4700~\AA~ between days 0--7 and 35--40, which is typically identified as the Fe \textsc{ii} triplet $\lambda$$\lambda$$\lambda$4924, 5018, 5169 (multiplet 42;  \citetalias{Maund07, Maund09, Tanaka12, Reilly16}).  At days 0--7, the polarization peak of this feature is blueshifted by $v = -17,630 \pm 510$  km s$^{-1}$ from the central Fe \textsc{ii} rest wavelength of 5018 \AA, as shown in Figure \ref{fig:e1_polar_plot}. This feature reaches a maximum polarization of $p = 2.29 \% \pm 0.34\%$ at $v = -12,850 \pm 440$  km s$^{-1}$ during days 57--67 and maintains a similar magnitude throughout the photospheric phase. 

At days 35--40 and 57--67, the region bluewards of the triplet ($v = -18,000$ to  $-14,000$ km s$^{-1}$) creates a loop in the first quadrant of the $q-u$ plane. At days 35--40 the PA over this loop signature rotates $> 45\degr$. By days 57--67 the PA rotates over a bigger angle, from $\approx200\degr$ to $\approx110\degr$ across the line (Fig. \ref{fig:polar_plots}). After days 57--67, the Fe \textsc{ii} triplet is no longer identifiable in the polarization spectrum. 

\subsection{Oxygen}\label{subsec: Oxygen}
The spectral region around O \textsc{i} $\lambda$7774 shows an increase in polarization above the continuum level throughout the photospheric phase (Fig. \ref{fig:specpol}). Its magnitude increases with time, reaching a maximum value of $p = 4.25 \% \pm 0.90\%$ for $v = -7,100 \pm 280$ km s$^{-1}$ at days 85--90. There is also an atmospheric absorption line in this region that we assume does not contribute any polarization. Although forbidden lines are not necessarily expected to be seen at early times, our spectra show a polarization peak at days 0--7 in the vicinity of the collisionally excited [O \textsc{i}] $\lambda\lambda$6300, 6330 doublet. During days 0--7, this peak is blueshifted from rest by the same amount as O \textsc{i} $\lambda$7774  ($v = -9,520 \pm 340$ km s$^{-1}$), and the two peaks share a similar polarization magnitude ($p = 0.95 \% \pm 0.05\%$ for the possible [O \textsc{i}] feature and $p = 0.85 \% \pm 0.05\%$ for O \textsc{i} $\lambda$7774).

At early times (days 0--7 and 35--40), the PA for the peak polarization value of  O \textsc{i} $\lambda$7774 agrees well with the GCF (\S~\ref{sec: continuum};  Fig. \ref{fig:polar_plots}), although it rotates ($\approx 30\degr$ in each case) across the line feature.  At days 0--7, the PA of the possible [O \textsc{i}] $\lambda\lambda$6300, 6330 peak polarization feature is the same as that of O \textsc{i} $\lambda$7774 ($\approx210\degr$). However, the PA across the entire [O \textsc{i}] feature roughly maintains this angle, rather than rotating across the line like the PA of O \textsc{i} $\lambda$7774. Although the two line regions align well in this earliest photospheric epoch (days 0--7), 
in later epochs it becomes harder to discern a well-defined polarization signature near [O \textsc{i}] $\lambda\lambda$6300, 6330.
Despite the coincidences with the O \textsc{i} $\lambda$7774 line behavior, it is possible this early-time polarization signature could be due to H$\alpha$/Si \textsc{ii} instead.

In the two later photospheric observations (days 57--67 and 85--90), a feature emerges at 6090 \AA~that is blueshifted from [O \textsc{i}] $\lambda\lambda$ 6300, 6330 \AA~by 
$v = -9990 \pm 340$ km s$^{-1}$. At this time the double-peaked profile of the [O \textsc{i}] $\lambda\lambda$ 6300, 6330 \AA~doublet starts to become apparent in the flux spectrum (\S~\ref{sec: spectral_evolution}). The consistency of this polarization signature through to the transitional-nebular phase (\S~\ref{subsec: epoch6 polarization}) suggests it is associated with [O \textsc{i}] $\lambda\lambda$ 6300, 6330\AA. However, the PA across this signature has large uncertainties for days 57--67 and does not agree with the PA values of O \textsc{i} $\lambda$7774 at days 85--90 (Fig. \ref{fig:polar_plots}).  

\subsection{Transitional-Nebular Phase Polarization }\label{subsec: epoch6 polarization}
Our final epoch at day 295 captures a uniquely transitional spectrum in both flux (\S~\ref{sec: spectral_evolution}) and polarization, revealing characteristics of the ejecta as it evolves into the transitional-nebular phase. As discussed above, we have not identified or subtracted the continuum (or ISP) from this epoch of data or any of the values quoted in this section (\S~\ref{sec: continuum}). The most distinct feature of this epoch is that the the bulk of the data forms a dominant axis newly directed towards the fourth quadrant of the $q-u$ plane (Figure \ref{fig:all_QU}). The GCF that defines this new dominant axis represents a clear rotation from $\theta = 12.0-24.1\degr \pm 1.9\degr$ in the first five photospheric epochs to $\theta =168.9\degr \pm2.3\degr$ in this transitional-nebular phase, a change over time that is independent of the ISP and provides insight into the continuum behavior (\S~\ref{sec: continuum}, Table \ref{Tab:continuum_q-u_bfit}). Furthermore, it is clear that this spectrum is largely polarized beyond our ISP upper limit and many of the continuum estimates from previous epochs, suggesting that a  
new scattering structure within the SN is visible at this time. Here we present the transitional-nebular phase polarization data along with a basic analysis, but we save the possible interpretations to be discussed in \S~\ref{subsec: disc_structure_over_time} and \S~\ref{subsec: disc_mechanism}. 

Of the several distinct emission features in the flux spectrum at day 295 (\S~\ref{subsec:nebular_flux}), none show strong changes in polarization aligned with their rest wavelengths as would be expected from a slowing ejecta (Figure \ref{fig:specpol}). One such signature is a broad polarization feature from 6500--7100 \AA~with a distinctly large magnitude ($p = 4.59 \% \pm 0.26\%$) compared to other polarization signatures seen in SN~2012au so far; this feature plays a significant role in the wavelength dependence of the data along this epoch's dominant axis (Fig. \ref{fig:all_QU}). However, after much investigation we cannot narrow down the origin of this feature beyond suggesting there are multiple potential contributing sources (including unseen spectral lines). 

\begin{table}
 \begin{tabular}{c c c c c}
    \hline
    Region & ${\overline{p}}~(\%)$ & $\sigma_{\overline{p}}~(\%)$ & $\overline{\theta}$ (\degr) & $\sigma_{\overline{\theta}}$ (\degr) \\
    \hline
    \multicolumn{5}{c}{Mg \textsc{i}] $\lambda$4571~\AA}\\ 
    \hline
    blue & 0.74 & 0.14 & 31.3 & 5.1\\
    central & 0.41 & 0.07 & 28.8 & 4.7\\
    red & 0.79 & 0.09 & 3.5 & 2.7 \\
    \hline
    \multicolumn{5}{c}{Fe \textsc{ii}] $\lambda\lambda\lambda$4924, 5018, 5169~\AA}\\ 
    \hline
    blue & 0.44 & 0.09 & 17.5 & 5.9 \\
    central & 0.60 & 0.08 & 6.8 & 3.4 \\
    red & 0.47 & 0.08 & 36.3 & 4.1 \\
    \hline
    \multicolumn{5}{c}{Na \textsc{i} D $\lambda\lambda$5890, 5896~\AA}\\ 
    \hline
    blue & 1.96 & 0.15 & $-$7.1 (172.9) & 2.2 \\
    central & 0.85 & 0.13 & 3.7 & 3.4 \\
    red & 1.64\tablenotemark{*} & 0.08 & $-$4.0 (176.0) & 1.9 \\
    \hline
    \multicolumn{5}{c}{[O \textsc{i}] $\lambda\lambda$6300, 6330~\AA}\\    
    \hline
    blue & 1.59\tablenotemark{*} & 0.08 & $-$3.3 (176.7) & 1.9 \\
    central & 0.62 & 0.05 & 10.9 & 2.1 \\
    red & 1.30 & 0.09 & $-$6.1 (173.9) & 2.0 \\
    \hline
 \end{tabular}
 \caption{Error-weighted mean polarization and PA values near the four major flux signatures seen in the day 296 transitional-nebular spectrum. We used three velocity regions for each line: a highly \textit{blue}-shifted region from -14000 to -5000 km s$^{-1}$, a symmetric low(er)-velocity \textit{central} region, -5000 to +5000 km s$^{-1}$, and a highly \textit{red}-shifted region from +5000 to +14000 km s$^{-1}$ (\S~\ref{subsec: epoch6 polarization}). Values in this table have not been continuum subtracted.}
 \tablenotetext{*}{Similar values due to the $\approx5000$ km s$^{-1}$ overlap between the two region.}
 \label{Tab:Nebular_pol}
\end{table}

\begin{figure*}
    \centering
    \begin{minipage}[b]{0.45\textwidth}
    \includegraphics[width=\columnwidth]{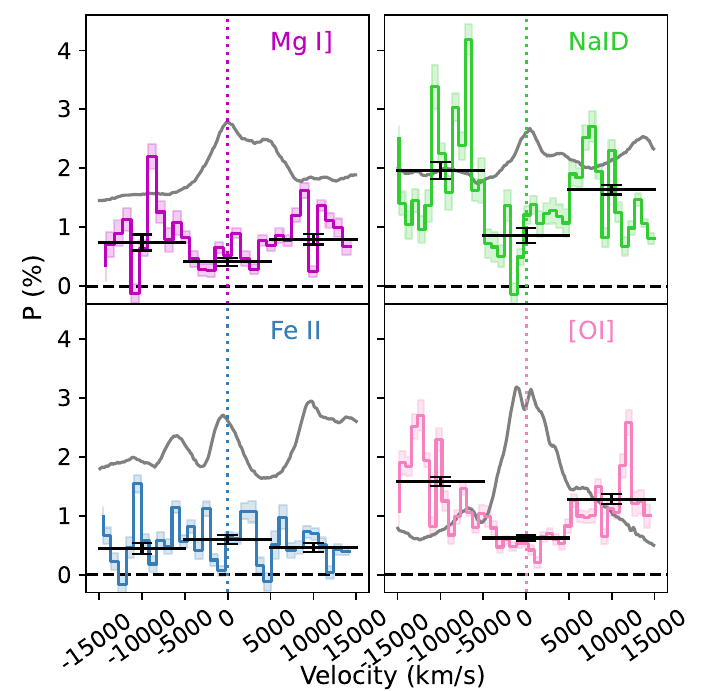}
    \caption{Day 295 polarization regions in velocity space near the rest wavelengths of major spectral lines (\S~\ref{subsec: epoch6 polarization}). The flux spectrum for each region is plotted in grey, while the corresponding region of each polarization spectrum is color coded and labeled by species. In each panel, the rest velocity of each line is plotted as a vertical dotted line color coded for that species, and $p=0$ is shown as a horizontal dashed black line. The error-weighted mean values for three velocity regions defined in \S~\ref{subsec: epoch6 polarization} are shown as black horizontal bars; these values are presented in Table \ref{Tab:Nebular_pol}. These data have not been continuum subtracted.}
    \label{fig:epoch6_Pspec}
    \end{minipage}
    \hfill
    \begin{minipage}[b]{0.45\textwidth}
    \includegraphics[width=\columnwidth]{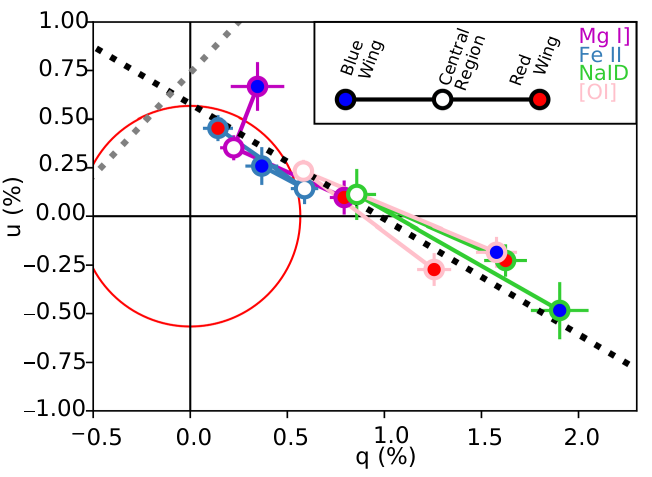}
    \caption{The same error-weighted mean polarization values shown for the transitional-nebular line regions in 
    Fig.~\ref{fig:epoch6_Pspec} and tabulated in Table \ref{Tab:Nebular_pol}, now displayed in $q-u$ space. Blue, white, and red points represent the blue wing (-15000 to -5000 km s$^{-1}$), central 
    region (-5000 to +5000 km s$^{-1}$), and red wing (+5000 to +15000 km s$^{-1}$) values, respectively. The GCF slopes for days 0--7 and day 295 are plotted as grey and black dotted lines, respectively; the Serkowski reddening maximum ISP estimate is displayed as a red circle. These data have not been continuum subtracted.}
    \label{fig:epoch6_qu}
    \end{minipage}
\end{figure*}

Although there are other polarization features in the spectra with $p\gtrsim2\%$, we focus our analysis of this epoch on the polarization near the four most pronounced emission features in the flux spectrum pertaining to the heavier elements: Mg \textsc{i} $\lambda$4571, Fe \textsc{ii} $\lambda\lambda\lambda$4924, 5018, 5169, Na \textsc{i} D $\lambda\lambda$5890, 5896 and [O \textsc{i}] $\lambda\lambda$6300, 6330. (At this late time, we attribute the polarization in the $\approx5800$ \AA~region to Na \textsc{i} D $\lambda\lambda$5890, 5896 as opposed to He \textsc{i} $\lambda$5876, which dominates during the photospheric phase, due to the nature of the line's flux profile and the difference in its polarization from the earlier observations (\S~\ref{subsec: Helium})). We defined three regions of velocity space surrounding each of these spectral lines: a high-velocity wing on either side of the rest wavelength (-14000 to -5000 km s$^{-1}$ and +5000 to +14000 km s$^{-1}$) and a central ``rest'' region symmetrically spanning lower velocities (-5000 to +5000 km s$^{-1}$). We then calculated the error-weighted mean values (from our unbinned data, \S~\ref{sec:Obs&reduction}) for each region and display these in Figure \ref{fig:epoch6_Pspec}, overlaid on the data for each line region in velocity space (binned to 15 \AA). We also list these mean values in Table \ref{Tab:Nebular_pol}. Additionally, in Figure \ref{fig:epoch6_qu} we show each of these mean values in the $q-u$ plane along with the GCFs for this epoch and days 0--7. 

The lines more commonly associated with transitional-nebular spectra (Mg \textsc{i} $\lambda4571$, Na \textsc{i} D $\lambda\lambda$5890, 5896, and [O \textsc{i}] $\lambda\lambda$6300, 6330) exhibit symmetrically mirrored polarization features about their rest wavelengths (although we note that with our definitions, the red wing of the Na \textsc{i} D line and the blue wing of the [O \textsc{i}] line overlap by $\approx$ 5000 km s$^{-1}$). For each of these lines, the high-velocity wing regions show an increase in polarization compared to the central ``rest'' region,
which is located closer to the origin in $q-u$ space along the line of the GCF (Figures \ref{fig:epoch6_Pspec} and \ref{fig:epoch6_qu}). Comparing the range of polarization magnitudes in these wing regions ($0.74\% \pm 0.14\% \leq p \leq 1.96\% \pm 0.15\%$) to the upper limit ISP estimate, it is apparent that at least some of the level of polarization seen in these regions is intrinsic to SN~2012au. 

Changes in the polarization across the transitional-nebular lines could be interpreted as arising from either the line scattering regions or the continuum-emitting surface at this late time, depending on one's interpretation of the continuum polarization. If the continuum level proceeded to decrease according to the trend established throughout the photospheric observations (Figure \ref{fig:cont_est}), then the lower polarization seen in the central line regions would be closer in magnitude to the expected levels at this time, compared to the high-velocity wing regions (Tables \ref{Tab:continuum_regions} and \ref{Tab:Nebular_pol}). In this case, there is a clear velocity offset between the line polarization signature in the wing regions and its rest wavelength (Figure \ref{fig:epoch6_Pspec}). Alternatively, it is possible that the central line regions are depolarized with respect to the surrounding continuum polarization, which is denoted by the high-velocity wing regions. In this scenario, the continuum polarization measured in the wings of Na \textsc{i} D $\lambda\lambda$5890, 5896 and [O \textsc{i}] $\lambda\lambda$6300, 6330 shows an increase between the last photospheric phase (days 85--90) and this transitional phase (day 295; Tables \ref{Tab:continuum_regions} and \ref{Tab:Nebular_pol}). The polarization in the wings of Mg I $\lambda$4571 is similar to that measured during days 85--90, suggesting a slight wavelength dependence to the continuum polarization at this time as shown by the location of the mean values with respect to each other in the $q-u$ plane
(Figure \ref{fig:epoch6_qu}).

On the other hand, the polarization surrounding the Fe \textsc{ii} triplet deviates from trends seen around the other lines, as it maintains a nearly constant magnitude ($p\approx0.55\%$) across all three regions in velocity space, while the central region mean is further from the origin than the high-velocity wing mean values in the $q-u$ plane. The mean values across all the iron regions are lower than the upper-limit ISP estimate and previous continuum estimates, indicating that the Fe \textsc{ii} triplet may not be a significant source for the polarization in this region.

\section{Discussion}\label{sec: discussion} 
It is clear from the nonzero continuum and line polarization signatures seen throughout the multi-epoch SP data we present here that SN~2012au exhibits significant departures from circular symmetry on the plane of the sky. The variations in magnitude and PA both within a single epoch and over time suggests that multiple asymmetric substructures are revealed over the course of the SN's evolution. Understanding the nature of these asphericities can provide key insights into the mechanism behind such a luminous explosion, as well as links to similar objects.

\citet{Stevance19} summarized the physics generally applied in interpreting SP observations. In our observations of SN~2012au we find all three classical SP signatures that these authors discussed:
\begin{enumerate}
    \item{Nonzero continuum  polarization caused by a globally aspherical distribution of electrons, e.g., a prolate or oblate ellipsoidal photosphere;} 
    \item{Increases (decreases\footnote{Depolarization can also arise from optically thin line emission outside the SN~photosphere
    even if the emitting material is not asymmetric, because of the additional unpolarized flux it may contribute.}) in polarization above (below) the continuum level, coinciding with specific line transitions, signifying an incomplete blocking of the photosphere by an uneven distribution of material; and}
    \item{Changes in polarization at later times after the photosphere has receded, implying the presence of energy sources such as (stalled) jets or off-center radioactive decay that create asymmetric illumination of the remaining material.}
\end{enumerate}

We use these physical cases, along with the \citet{WW08} SP classification system, to compare our findings for SN~2012au with the results for similarly studied (using SP data) SNe Type Ib and comment on the characteristics of this class of SNe as a whole. We then describe the possible explosion geometry of SN~2012au at days 0--7, 26, 35--40 and 295. Finally, based on our comparisons to other SNe and proposed geometries, we discuss the potential mechanisms responsible for SN~2012au's explosion and revealed structures. 

\subsection{Comparison of SNe Ib Explosion Structures}\label{subsec: disc_comp_other_Ibs}

\begin{table*}
\centering
\begin{tabular}{|c|c c c c c|}
\hline
    & SN 2005bf & SN 2008d & SN 2009jf & iPTF 13bvn & SN 2012au \\
    \hline
    Epochs Observed & 2 & 2 & 1 & 6 & 6 \\
    \hline
    Comparable Epoch (day) & 8 (D0) & 3.3 (N1) & 9.3 (N1) & 7 & 0-7 (D0)\\
    $\%p_\textrm{cont}$  & 0.45--1.2 & 0.22$\pm$0.14 & -- & 0.30$\pm$0.13 & 0.98$\pm$0.05\\
    Photosphere asymm. ($\%$)& 20--50 & 10 & -- & 10--15 & 30 \\
    $\%p_\textrm{{Na \textsc{i}/He  \textsc{i}}}$ & 0.6$\pm$0.4 & 0.4$\pm$0.2 & 0.5$\pm$0.2 & 0.9*& 1.40$\pm$0.05 \\
    $\%p_\textrm{Fe \textsc{ii}}$ & 0.4$\pm$0.2& 0.8$\pm$0.2& 0.4$\pm$0.2& 0.6* &2.23$\pm$0.05\\
    $\%p_\textrm{O \textsc{i}}$ & 0.0$\pm$0.5 & 0.5$\pm$0.13 & 0.9$\pm$0.2 (L)& -- & 0.85$\pm$0.05\\
    $\%p_\textrm{Ca \textsc{ii}}$ & 1.5$\pm$0.3 & 1.8$\pm$0.3 &  1.2$\pm$0.2 (L)& $\sim$2.5 (L) & --\\
    \hline
    Other Epoch (day) & -6 (D0) & 18.3 (N1) &  -- & 25 & 26 (L) \\
    $\%p_\textrm{cont}$ & 0.45 & 0.21$\pm$0.17 & -- &  0.3$\pm$0.13 & 1.17$\pm$0.05\\
    Photosphere Asymm. ($\%$)& $\gtrsim$10 & $>$10 & -- & $\sim$10--15 & 10--40 \\ 
    $\%p_\textrm{Na \textsc{i}/He  \textsc{i}}$ & 1.3 (L) & 0.68 (L) & -- & 0.8* (L) & 2.58$\pm$0.05 (L) \\
    $\%p_\textrm{Fe \textsc{ii}}$ & $\sim$1.5 (L) & 1$\pm$0.3 & -- & 0.5* & 1.93$\pm$0.05 (L) \\
    $\%p_\textrm{O \textsc{i}}$ & $\sim$0 & 0.3$\pm$0.13 & -- & -- & -- \\
    $\%p_\textrm{Ca \textsc{ii}}$ & 4$\pm$1 (L) & 2.5$\pm$0.7 (L) & -- & 3.3$\pm$0.8 & --\\
    \hline
    Reference & \citetalias{Tanaka09} (day 8), \citetalias{Maund07} (day -6) & \citetalias{Maund09} & \citetalias{Tanaka12} & \citetalias{Reilly16} & This paper\\
    \hline 
\end{tabular}
\caption{Polarization in the continuum and Na \textsc{i}/He \textsc{i} $\lambda$5876, Fe \textsc{ii} $\lambda$5018, O \textsc{i}  $\lambda$7774 and Ca \textsc{ii} $\lambda$8498 features for all Type Ib SNe SP data published thus far. Each SN~has a different number of observations, but we list polarization properties for each at a comparable epoch (in days since $R$-band maximum) to our first observation of SN 2012au.
The ellipticity for the photospheres are inferred from \citet{Hoflich91} and quoted as a percent. Continuum values quoted for SN 2012au are averages of the blue and red continuum regions listed in Table \ref{Tab:continuum_regions}. Line polarization has been continuum subtracted. Values are quoted directly from the references listed (some with uncertainties and others as approximations) or from \citetalias{Tanaka12};  we denote with an asterisk values we inferred from figures in these references. The SP type we assigned to each spectrum is labeled next to the epoch, and we list SP type L next to individual line polarization values for lines that exhibited clear loops in the $q-u$ plane.}
\label{Tab:Ib_SNe}
\end{table*}

\begin{figure}
    \centering
    \includegraphics[width=1.25\columnwidth]{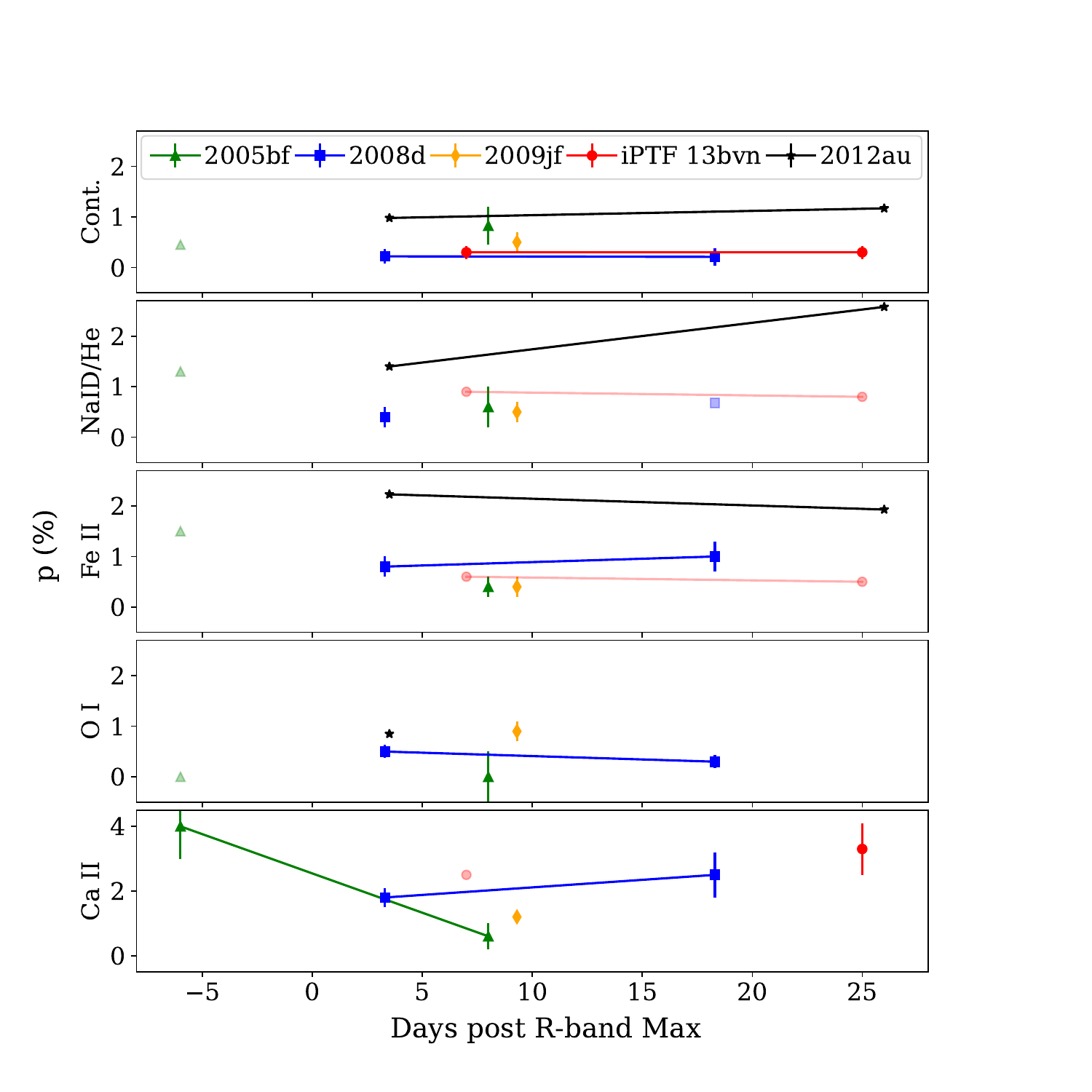}
    \caption{Comparison of polarization measurements by line species (or continuum) and days post $R$-band maximum  for all Type Ib SNe SP data published thus far. Values shown are quoted in Table \ref{Tab:Ib_SNe} with citations. Each SN is specified by a different color and marker style. In many cases the level of uncertainty is indistinguishable on the points plotted, so  measurements with unknown uncertainties are displayed as lighter colored points.}
    \label{fig:table6_compSNe}
\end{figure}

Here we compare the polarization trends seen for all SNe Type Ib with published and analyzed SP data to date: SN~2005bf (\citetalias{Maund07, Tanaka09}), SN~2008d (\citetalias{Maund09}), SN~2009jf (\citetalias{Tanaka12}), SN~iPTF 13bvn (\citetalias{Reilly16}), and SN~2012au. In the top half of Table \ref{Tab:Ib_SNe} we list continuum and line polarization values for comparable early epochs (defined by days post $R$-band maximum) for each of these targets, providing as much information on the measurements as was available from each study. In the bottom half of the table we list these values for the next most interesting epoch of data available for each object (SN 2009jf was only observed for one epoch). All studies applied some form of ISP and/or continuum subtraction, so all values quoted in this comparison represent intrinsic polarization values unless otherwise stated. The polarization values quoted in Table~\ref{Tab:Ib_SNe} are also displayed in Figure \ref{fig:table6_compSNe}. Additionally, in Table \ref{Tab:Ib_SNe} we assign the SP type classification \citep{WW08} for each epoch listed, and note when a specific line region created a loop by adding the SP type (L) next to the line polarization value. 

Within this sample of SNe Ib studied with SP, SN~2005bf and SN~2012au stand out as unique objects. The continuum polarization for SNe 2008d, and iPTF 13bvn ranges from $p = 0.21\%-0.30\%$, and estimates for the different epochs observed for each object were in agreement with each other, showing that there was little variation in the continuum polarization of these objects over the course of the reported observations. In comparison, continuum levels for SN~2005bf ranged from $p = 0.45\%-1.2\%$, while for SN~2012au we found $p = 0.32\%-1.47\%$ (without ISP subtraction; Table \ref{Tab:continuum_regions}), and both SNe showed a general decrease in continuum polarization throughout the photospheric phase (\citetalias{Tanaka09}). These measurements suggest that at least some SNe Ib possess elliptical photospheres; we compare their inferred ellipticities in Table \ref{Tab:Ib_SNe} \citep{Hoflich91}. 

The decrease in continuum polarization seen during the photospheric phase in SN~2012au and SN~2005bf could be due either to the photosphere becoming more symmetric, to increasing polarimetric cancellation, or to the optical depth decreasing due to the homologous expansion of the atmosphere, resulting in fewer scattering events overall \citep{Tanaka09}. While the latter  explanation is more commonly associated with the transitions of SNe II to the transitional-nebular phase (when the outer hydrogen envelope becomes more transparent; \citealt{Leonard06, Dessart11, Dessart21b}), the principle also holds for SNe Ib. As the ejecta dissipate, the photosphere recedes in mass coordinates, revealing the geometry of the inner explosion. Thus a decreasing trend in the continuum polarization
may be due to the outer ejecta becoming transparent, revealing a photosphere that has receded to a more symmetric inner layer.\footnote{We emphasize that this interpretation refers only to the observations of SN~2012au during days 0--90 post-maximum and that due to its slowly evolving nature, the inner layer we invoke at this point is most likely not the innermost core structure. We discuss the SN~core in \S~\ref{subsec: disc_structure_over_time} and \S~\ref{subsec: disc_mechanism} \citep[see also][]{MiliD13, Pandey21}.}

Additionally, both SN~2005bf and SN~2012au exhibit dominant axes in the $q-u$ plane (SP type D0) at early times, with the polarization magnitudes having a slight inverse wavelength dependence (\citetalias{Tanaka09}). The presence of this dominant axis (which we quantify as the GCF for SN~2012au; \S~\ref{sec: continuum}), along with these objects' relatively high continuum polarization, suggests that the photospheres of SN~2005bf and SN~2012au were more elongated compared to the others at early times (Table~\ref{Tab:Ib_SNe}). This apparent greater elongation could either be intrinsic or the result of a viewing angle effect that resulted in SN~2005bf and SN~2012au being oriented at more favorable angles to reveal their asymmetric structure.

All five of the Type Ib SP sample objects exhibit loops in the $q-u$ plane for multiple line regions during the early photospheric phases (observations range from days -6 to 40); however, these loops do not appear to persist for more than $\approx$10 days. Most commonly noted are loops across He \textsc{i} $\lambda$5876, which are present in the spectra of all but SN~2009jf, as well as those seen in the Ca \textsc{ii} IR triplet region (which unfortunately is beyond the spectral range of our SN~2012au observations). \citetalias{Maund09} also mentions less pronounced loops for the He \textsc{i} $\lambda$6678 and $\lambda$7065 lines of SN~2008d, which we also see in SN~2012au at days 26 and 35--40 (Fig. \ref{fig:He_lines_QU_vel}). These loops represent changes in polarization and PA away from the continuum, suggesting that the majority of the SN Ib sample objects exhibit incomplete blocking of the photosphere and thus that their explosions contain possess complex, multi-component structures \citep{WW08}. 

Most of the observed loops span a 100--300 \AA~ range (6000--10000 km s$^{-1}$) located somewhere between $v=0$ and $v = -15000$ km s$^{-1}$ \citep{WW08}. However, the helium loops in SN~2012au span velocities from 0 to -18000 km s$^{-1}$ (Fig. \ref{fig:He_lines_QU_vel}), supporting the picture that this explosion was more energetic than the others and propelled material to higher velocities \citep{Pandey21}. The peak polarization levels for the He \textsc{i} $\lambda$5876 loops in SN~2012au are also $0.5\%-1.9\%$ higher than those seen at comparable times in the other SNe (Table \ref{Tab:Ib_SNe}). We discuss an interpretation for SN~2012au's helium loops further in the next section (\S~\ref{subsec: disc_structure_over_time}). 

For SN~2005bf and SN~2008d, the epochs when the loops are seen also correspond to the single epoch (day -6 and day +18.3 respectively) when the PAs of the He \textsc{i} lines ($\lambda$5876, $\lambda$6678, and $\lambda$7065) and Fe \textsc{ii} $\lambda$5018 differ by $\approx$45\degr  \citep{Maund07, Maund09}. SN~iPTF 13bvn exhibits a $\approx$90\degr difference between He \textsc{i} $\lambda$5876 and Fe \textsc{ii} $\lambda$5018 for a single observation (day +9), although \citetalias{Reilly16} mention no signs of helium loops at this time. At all other epochs for these SNe and for SN~2009jf, the polarization signatures for the helium and iron lines show similar behavior to each other, suggesting the distributions of the two materials are correlated  \citep{Maund07, Maund09, Tanaka12, Reilly16}. 

For SN~2008d, \citetalias{Maund09} suggested that when the PAs and velocities of the two materials agreed (day +3.3), the Fe \textsc{ii} line region may have been blended with  He \textsc{i} $\lambda$5015, while later on when their polarization properties differ (day +18.3), He \textsc{i} $\lambda$5876 may have been blended with Na \textsc{i} D $\lambda\lambda$5890, 5896, creating the observed loop. \citetalias{Reilly16} proposed that the dual-velocity structure they observed across He \textsc{i} $\lambda$5876 in SN~iPTF 13bvn is similarly due to blending with Na \textsc{i} D $\lambda\lambda$5890, 5896. SN~2012au is unique in that the distinct difference in helium and iron behavior persists for all three of the He \textsc{i} lines over the course of multiple epochs, from days 0--40 (Fig. \ref{fig:polar_plots}, Fig. \ref{fig:He_lines_QU_vel}). Furthermore, SN~2012au shows no sign of either Fe \textsc{ii} $\lambda$5018 being blended with  He \textsc{i} $\lambda$5015, or  He \textsc{i} $\lambda$5876 being blended with  Na \textsc{i} D $\lambda\lambda$5890, 5896 over the course of days 0--40 (\S~\ref{subsec: Helium}). 


In terms of orientation, the loops seen in SN~2005bf are most similar to those of SN~2012au, as they too are oriented orthogonal to the dominant axis in the $q-u$ plane (\citetalias{Maund07, Tanaka09}). The differences between the PAs of the most extended points in the He \textsc{i} $\lambda$5876 loops and the dominant axis for SN~2005bf (day -6) and SN~2012au (day 26) are $\approx$32.3\degr and $\approx$67.5\degr, respectively (Tables \ref{Tab:continuum_q-u_bfit} and \ref{Tab:phot_line_pol}). This implies that in both SNe, the He \textsc{i} line-scattering region is oriented differently than the photosphere. \citetalias{Tanaka09} suggested that the difference between the elements creating the $q-u$ loops and the rest of the ejecta material is due to an uneven $^{56}$Ni distribution creating selectively excited or asymmetrically illuminated regions. In their illustration of SN~2005bf, they depicted this uneven distribution as ``flared'' He, Si, and Ca regions atop an underlying $^{56}$Ni-rich, unipolar or bipolar cone-shaped outflow at an angle offset from the elongation axes of both the photosphere and the inner core.

Although none of the other SNe were observed using SP at late enough times to capture the transition to the transitional-nebular phase (when signatures of the inner core are more visible;  \citealt{Leonard06, Dessart11}), each of the Ib SP studies reports line polarization (or depolarization) in signatures of heavier elements and uses these to make a connection to the geometry of the inner ejecta. \citetalias{Maund07}, \citetalias{Tanaka09}, \citetalias{Maund09}, and \citetalias{Reilly16} conclude that if each of these SNe Ib were caused by a jet explosion mechanism, the jet must have stalled in the core, based on the low Fe \textsc{ii}, O \textsc{i} and continuum polarization levels (Table \ref{Tab:Ib_SNe}). While \citetalias{Tanaka12} do not explicitly state these same conclusions, we deduce the same is true for SN~2009jf based on their report of similar low polarization in Fe \textsc{ii} and O \textsc{i}. 

In the context of a stalled jet, larger line polarization values could imply that the jet is nearly aligned with our line of sight and the line polarization we observe in SN 2012au is produced by multiple scattering events within the jet. Alternatively, larger line polarization could be due to our viewing the ejecta at an angle close to orthogonal to the jet axis. In order to compare our findings with previously published models of such systems (which are simplified to compute polarization for single scattering only; \citealt{Tanaka17}) we focus our discussion on the latter of these two scenarios. 

The two largest polarization values reported for the Fe \textsc{ii} lines in SNe Ib are in the earliest epochs of SN~2005bf (day -6, $p = 1.5\%$; \citetalias{Maund07}) and SN~2012au (days 0--7, $p = 2.23\%$). However, the polarization levels observed for iron and oxygen lines in SN~2012au are greater than those reported for the other objects at the nearest comparable epoch by $\Delta p = 1.43\%$ for iron and $\Delta p = 0.35\%$ for oxygen (Table \ref{Tab:Ib_SNe}; the only exception is a similarly polarized oxygen line in SN~2009jf). If we take the explosions SN~2012au and SN~2005bf to be at similar optimum viewing angles, then the continually larger line polarization and mostly larger continuum polarization  of SN~2012au (Table \ref{Tab:Ib_SNe}) suggest that the geometrical structure of SN~2012au was more asymmetric than was SN~2005bf. 

In the next two subsections we present a picture of the possible explosion geometry of SN~2012au (\S~\ref{subsec: disc_structure_over_time}) and discuss the powering mechanism (\S~\ref{subsec: disc_mechanism}), while referencing the comparisons made here to the other type Ib SNe and particularly to SN~2005bf. While we do not delve into the details of the SNe Type IIb studied with SP, we note that SN~1993J and SN~2011dh also exhibited early time dominant axes and loops coinciding with the He \textsc{i} $\lambda$ 5876 line, suggesting similar physical structures exist in the close neighboring SN types \citep{Mauerhan15, Stevance20}. 

\subsection{Time-Dependent Asymmetric Structure}\label{subsec: disc_structure_over_time}

We have identified four distinct stages in the ejecta development of SN~2012au. In Stage I (days 0--7), we classify SN~2012au as SP type D0 with a dominant axis that results in GCF$\theta = 23.3\degr$ (\S~\ref{sec: continuum}, Table \ref{Tab:continuum_q-u_bfit}). The distribution of data exhibits a slight wavelength dependence along this dominant axis (Fig.~\ref{fig:all_QU}). At this time, a high  continuum polarization indicates a large-scale aspherical geometry for the ejecta, such as an elongated ellipse deviating from spherical symmetry by greater than 30$\%$ \citep[Table~\ref{Tab:continuum_regions};][]{Hoflich91}. In this stage the bulk of the ejected material shares the same PA as the GCF (creating this dominant axis in the $q-u$ plane), including the line polarization associated with the Fe \textsc{ii} triplet, which creates a distinct inverse P Cygni profile in the polarization spectrum (Fig. \ref{fig:specpol}). However, the helium-rich material differs slightly in PA, with the $\lambda$5876 line diverging from the dominant axis (Fig. \ref{fig:polar_plots}) and creating a loop in the $q-u$ plane (\S~\ref{subsec: Helium}; Fig.~\ref{fig:He_lines_QU_vel}).    

\begin{figure}
    \begin{minipage}[b]{0.40\textwidth}
    \centering
    \includegraphics[width=\columnwidth]{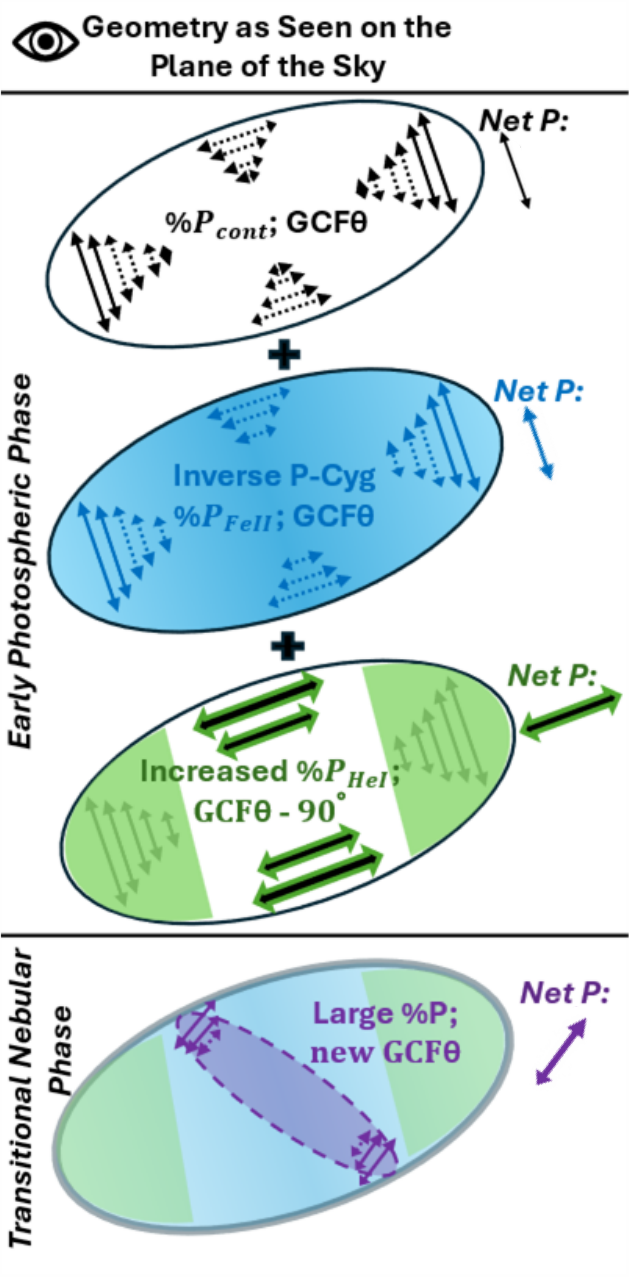}
    \caption{Illustration of the possible SN~2012au explosion geometry on the sky for days 0--40 (top three panels) and day 295 (bottom panel). Arrows depict polarization vectors, with dashed lines representing those that cancel out and solid lines representing those contributing to the net polarization (``Net P'' arrows). At early times, an elongated (prolate ellipsoidal) photosphere creates a net continuum polarization (\textit{top panel}). A nearly uniform distribution of iron over the photosphere (with a slight overdensity along the equatorial plane; \textit{second sketch}) creates a large polarization magnitude and inverse P Cygni polarization profile in the Fe \textsc{ii} triplet. Helium-rich material in the polar regions selectively blocks continuum light   such that the remaining continuum  creates the net polarization and rotated PA in the He \textsc{i} lines (\textit{third sketch}). At later times (\textit{bottom}), a highly asymmetric inner structure with a different PA reflects a newly oriented photosphere or excited heavy elements (purple). These sketches represent only one possible explanation for the observed signatures; other explosion configurations cannot be ruled out without further modeling.} 
    \label{fig:pos_geom}
    \end{minipage}
\end{figure}

In Stage II (day 26 and days 35--40), the polarization develops to reveal multiple components in the ejecta geometry, which we depict in the top three panels of Figure \ref{fig:pos_geom}. At this time the bulk of the polarization behaves similarly to Stage I. The dominant axis roughly maintains the same angle established at days 0--7, suggesting that the photosphere remains elongated; we depict this geometry as a prolate ellipsoid (Fig.~\ref{fig:pos_geom}) in accordance with the models of \citet{Hoflich91}. The distinct inverse P Cygni profile in the Fe \textsc{ii} triplet line polarization persists in Stage II, tracing a predominantly linear pattern in the $q-u$ plane at the same PA as the GCF (Fig. \ref{fig:He_lines_QU_vel}). As discussed in \citet{Doug01}, this behavior suggests a nearly uniform distribution of the iron-rich material across the same geometrical area as the photosphere. One caveat to this interpretation is the high polarization magnitude of the Fe \textsc{ii} triplet, which may indicate that a slightly higher density of iron along the equatorial plane of the photosphere creates enhanced polarization (Fig. \ref{fig:pos_geom}, second panel). In fact, 2D models invoking a bipolar asymmetry via overdense regions of iron placed either at the poles or in a toroidal geometry replicate a linear polarization signature in the $q-u$ plane \citep[\citetalias{Tanaka12};][]{Tanaka17}. Although this line shows a hint of a diverging $q-u$ loop at days 35--40 (Fig.~\ref{fig:He_lines_QU_vel}), we do not consider this significant due to the uncertainties on these data points.

The defining polarization features of this stage are the large He \textsc{i} loops in the $q-u$ plane, which span velocities from 0 to -18000 km s$^{-1}$ (Fig. \ref{fig:He_lines_QU_vel}). At day 26, we classify the explosion as SP type L because of these distinct loops, and note that at days 35--40 the loops have already begun to collapse to cover a smaller range of angles. Our measurement of the photospheric velocity at this time is $\approx$  -12500 km s$^{-1}$ (Fig. \ref{fig:polar_plots}); the difference between this velocity and that of the fastest part of the loop signatures corresponds to the distance the helium-rich region may extend beyond the photosphere. Since we are unable to see material beneath the photosphere, the portions of the loops seen at velocities slower than the photospheric velocity could simply represent the polarization returning to the continuum level, or may suggest that there is some viewing angle effect at play. 

\citet{Tanaka17} explored 3D models invoking varying clumpy geometries to reproduce loop signatures. In their models, an outer shell of clumpy ejecta absorbs line photons, creating a preferential, uneven cancellation of continuum polarization that produces loops in the $q-u$ plane. They find that the spread of the loops on the $q-u$ plane is a function of the clump sizes ($\alpha_{cl}$) and number of clumps (defined by a covering factor $\textit{f}_{cl}$; \citetalias{Tanaka12}; \citealt{Tanaka17}). Comparing the Doppler velocities and shapes from these models with the loops seen in SN~2012au might suggest that the helium is distributed among several large clumps over the entire face of the photosphere (0.2 $\leq$ $\textit{f}_{cl}$ $\leq$ 0.5 and $\alpha_{cl} \geq$ 0.5). However, the spherical photosphere underlying these models ignores the presence of the dominant axis and thus the degree of rotation the loops extend away from this. We propose that because the helium loops in SN~2012au exhibit a sharp rotation (nearly 90$\degr$) away from the dominant axis and produce polarization magnitudes well above the continuum estimate (and those in the Tanaka models), they are signs of a more nuanced geometry than those investigated by \citet{Tanaka17}. 

We surmise that the helium loops are caused by hot, helium-rich material located in the polar regions of the elongated photosphere, which selectively absorbs/emits at helium wavelengths (Fig. \ref{fig:pos_geom}). In this scenario, which is similar to the clumpy models in \citet{Tanaka17}, the increased polarization in the observed line is due to the (line-scattering) polar material unevenly blocking the photosphere. Keeping in mind that the polarization of the escaping light is oriented at a $90\degr$ angle from the plane of the blocking material, we use the PA of the peak He line polarization (which describes the overall divergence of these helium loops from the GCF) as a proxy for the orientation of the unblocked continuum region. The absorbing material in the polar regions thus creates a higher net polarization for the remaining continuum photons at the helium wavelengths, causing the increased polarization at the rotated PA observed in the helium lines. Furthermore, because this blocking material interacts only with helium line photons, it maintains the elongated appearance of the photosphere at all other wavelengths and thus roughly preserves the overall GCF$\theta$ shared between Stages I and II. 

In the previous section (\S~\ref{subsec: disc_comp_other_Ibs}) we noted that differences in the He \textsc{i} and Fe \textsc{ii} line signatures are common in most of the SP studied Type Ib SNe. Figures \ref{fig:e1_polar_plot} and \ref{fig:He_lines_QU_vel} best show the differences in polarization between the major He \textsc{i} and Fe \textsc{ii} lines seen in SN~2012au at these early stages. In our proposed scenario, the difference in the distribution of material may trace individual ionization temperatures such that helium is present in hotter regions and iron in cooler regions. Alternatively, it may be a sign that the lighter, hotter material (helium) was selectively expelled to the outer regions of the ejecta. In \S~\ref{subsec: disc_mechanism} we delve into the physical implications of these observational interpretations.

Stage III (days 57--67 and 85--90) represents a transitional phase between the early-time ejecta configuration and a distinctly different structure appearing in Stage IV. Due to the large scatter in the polarization seen in Figures \ref{fig:all_QU} and \ref{fig:polar_plots}, our observations at this stage reveal little about the geometry of the photosphere (\S~\ref{sec: continuum}) and line regions (\S~\ref{sec: line polarization}). 

In Stage IV (day 295), polarization levels (not clearly coinciding with specific lines) have increased to the maximum values observed in our dataset, and the bulk of the material creates a new dominant axis. The PA of the dominant axis at this stage has rotated from its orientation of $13.7\degr-23.3\degr\pm1.1\degr$ in Stages I and II to $168.9\degr \pm 2.3\degr$, as depicted in Figure \ref{fig:pos_geom}. This dramatic change in angle could be due to the transition from a prolate to an oblate ellipsoidal photosphere (\S~\ref{sec: continuum}) caused by the recession of the photosphere through different density structures \citep[][and references therein]{Stevance19}. However, the polarization signatures at this time only indicate an elongated structure; further modeling is needed to determine its specific shape. The depolarization at the line centers with respect to the continuum at this stage (Fig. \ref{fig:epoch6_Pspec}), when SN~2012au is beginning to enter a transitional-nebular phase, suggests there is still significant continuum polarization. 

In \S~\ref{subsec: epoch6 polarization} we 
discussed two different interpretations of these observations. If the lines (shown in Fig.~\ref{fig:epoch6_Pspec}) are truly depolarized at this time to the level seen in their central regions (near rest velocity) and the surrounding wing regions still represent the continuum, then we observe an increase in continuum polarization from Stage III to Stage IV  (\S~\ref{subsec: epoch6 polarization}). Since the continuum light and polarization arise from the photosphere, an increase in continuum polarization along with the change in the GCF$\theta$ tells us that the late-time photosphere has receded to reveal a newly oriented core structure (GCF$\theta=168.9\degr \pm 2.3\degr$), which is more asymmetric than the surface of the larger expanded photosphere from Stages I and II. However, if the line polarization is actually located in the high-velocity wing regions at this stage, then the polarization about these lines creates inverse P Cygni profiles. In this case, the offset of these polarization signatures from zero velocity suggests there is a significant separation between the location of the line emission and the scattering surface in the ejecta (Fig.~\ref{fig:epoch6_Pspec}). The common angle and level of polarization in each of the wings hint at a more axially symmetric ejecta, in which the material is more evenly distributed in an elongated structure on the plane of sky that is also oriented at the new GCF$\theta$ (Fig.~\ref{fig:epoch6_qu}). The mirrored nature of these wing-region polarization signatures about each line may reflect decreasing opacity in the ejecta, which enables us to observe both the receding and approaching ends of a bipolar structure.

At this late time, both the clear increase in the overall magnitude of polarization (regardless of whether we interpreted it as continuum or line polarization) compared to our earlier spectra and the presence of the newly oriented dominant axis point to a highly asymmetric inner structure. Furthermore, the PA of the new dominant axis, which is orthogonal to the PA of the high-velocity portions of the helium loops in day 26 (Fig.~\ref{fig:polar_plots}), is now offset from the orientation of the bulk of the material. Stages I and II suggests that any mechanism governing the inner structure also influenced the direction in which the outer, lighter material layers were ejected (Fig.~\ref{fig:pos_geom}).  

\subsection{Evidence for Explosion and Nebular-Phase Powering Mechanism}\label{subsec: disc_mechanism}
 
SN~2012au's categorization as a hypernova, superluminous SN, and magnetar host frames its explosion mechanism as a potentially integral connection among these CCSN subspecies \citep{Takaki, MiliD13, Pandey21, Omand23, Dessart24prep}. Our uniquely late SP observation of SN~2012au at day 295 probes the inner layers of the ejecta and provides geometrical information that may retain a direct imprint of the explosion mechanism. The overall larger polarization magnitudes seen for SN 2012au at day 295 compared both with earlier epochs and with other SNe Ib polarization measurements (\S~\ref{subsec: disc_comp_other_Ibs} \S~\ref{subsec: disc_structure_over_time}) indicate that this inner structure had a larger deviation from spherical symmetry than any previously observed SN Ib.

\citet{Omand23} used the transitional-nebular spectral lines present in SN~2012au to dissect this asymmetry. They suggested that lines with broader widths ($v_{FWHM} > 4500$ km s$^{-1}$) such as  Na \textsc{i} D $\lambda$$\lambda$5890, 5896 and [Ca  \textsc{ii}] $\lambda$$\lambda$7291, 7324 (in addition to [O \textsc{i}] $\lambda$$\lambda$6300, 6364 and Mg\textsc{i} $\lambda$4571) may arise from a different region of the ejecta than those with narrower widths ($v_{FWHM}\sim$2000 km s$^{-1}$) such as O \textsc{i} $\lambda$7774, O \textsc{i} $\lambda$1.317 $\mu$m and Mg \textsc{i} $\lambda$1.503 $\mu$m. Similar line widths were also reported in a survey of SLSNe by \citet{Gal-Yam16}, supporting the claim that SN~2012au is an intermediate case between this subclass and a typical SN~Ib. While our polarization spectra for day 295 do not encompass the latter (narrow width) emission lines for a full comparison, our analysis of the polarization potentially associated with the broader lines (except [Ca \textsc{ii}] $\lambda$$\lambda$ 7291, 7324) supports this claim: these features all share the same PA, suggesting they arise from a common scattering location (Figs. \ref{fig:polar_plots} and \ref{fig:epoch6_qu}). 

Additionally, the flux profiles of both the double-peaked [O \textsc{i}] $\lambda$$\lambda$6300, 6364 feature and the broadened Mg \textsc{i}] $\lambda$4571 feature are both shifted from rest by approximately -2500 km s$^{-1}$ (\S~\ref{subsec:nebular_flux}). \citet{Taubenberger2009} suggests that the asymmetric flux profile of Mg \textsc{i}] indicates clumping or a unipolar jet breaking the spherical symmetry of the ejecta. \citet{Maeda07} observed a similar double-peaked flux profile and velocity shift for [O  \textsc{i}] $\lambda\lambda$ 6330, 6364 in SN~20005bf during its transitional-nebular phase, which they attributed to the material being distributed in either an elongated structure directed towards the observer or a torus edge-on to the observer. Double-peaked [O \textsc{i}] lines produced in bipolar explosion models also support this picture \citep{Maeda08}.

It is clear from both the emission line profiles in SN 2012au and the high levels of polarization forming a dominant axis at our latest epoch (day 295) that the inner ejecta are highly asymmetric and have a preferred orientation (Fig.~\ref{fig:all_QU}). The PA rotation of the dominant axis between the early photospheric phase and the transitional-nebular phase demonstrates that the inner region of the ejecta is misaligned with respect to the initial direction of the bulk of the ejecta, creating a dual-axis structure (Table~\ref{Tab:continuum_q-u_bfit}; Fig.~\ref{fig:pos_geom}; ~\S~\ref{subsec: disc_structure_over_time}).

The level of asymmetry present in this dual-axis structure, as well as its evolution alongside the slowly evolving spectra of SN~2012au, may be connected to the efficiency of the explosion mechanism. \citet{Burrows21} reported that hydrodynamical models with longer delayed explosions yield a higher degree of asymmetry in the resulting CCSNe than do quickly exploding models. \citet{Maeda03} theorized that material accreting onto the center of the star over a longer time frame results in a less efficient explosion mechanism, which in turn produces a slow-evolving SN. Over the course of SN~2012au's slow evolution, it exhibited a dominant axis at both early and late times. The change in the PA of the dominant axis over time demonstrates that the material in the outer layers of the explosion was ejected in a direction orthogonal to that of the inner material. 

Since SN~2005bf also showed a slight misalignment between ejecta components at early times (\S~\ref{subsec: disc_comp_other_Ibs}) but was characterized to be a typical SN Ib, the more dramatic and persistent misalignment in SN~2012au could be a defining feature of slow-evolving, bright SNe. A dual axis at early times was also seen for the Type IIn SN 1997eg and attributed to the misalignment between the SN ejecta and the CSM it interacted with \citep{Hoffman08}. Additionally, similar large asymmetries to SN~2012au and a pronounced dominant axis were also observed in the early-time (day -23 and day +27) SP observations of the slow-evolving, superluminous Type Ic SN~2015bn \citep{Inserra16}. Though a dual axis has been observed at early times for multiple types of CCSNe, none of these studies included 
nebular phase observations to check whether these axes change over time as seen in SN~2012au. More SP observations of CCSNe near the transitional-nebular phase may help solidify a connection between the evolving ejecta structures and the efficiency of the explosion mechanism.

Both hypernovae and SNe hosting magnetars have also been found to possess multi-component structures with traits similar to those we identified for SN~2012au. \citet{{Maeda03}} found that the implementation of a high-velocity outer component and higher-density inner component was necessary for their models to accurately reproduce hypernova light curves. These models also explain the presence of broad absorption features like those seen in the early epochs (days 0--40) of SN~2012au (Fig.~\ref{fig:specpol}). Simulations of magnetohydrodynamical effects on CCSNe caused by a progenitor with magnetic field lines inclined to its axis of rotation resulted in a dual-component structure \citep{mikami08}. In these models, a magnetic torus forms around the inner proto-neutron star and jets are launched perpendicular to the torus \citep{mikami08}. These early magnetohydrodynamical models necessitate a progenitor star with a slightly stronger magnetic field than standard, which may in turn be the signature of a magnetar progenitor \citep{mikami08}. Since a magnetar has previously been suggested as the nebular-phase powering mechanism for SN~2012au \citep{MiliD18, Omand23, Dessart24prep}, the findings of \citet{mikami08} suggest that the misaligned structures we identified in its ejecta are to be expected.   
 
Several studies exploring models with magnetar powering sources have used SN~2012au for comparison because of its visibility at late times and its lack of CSM interaction \citep{MiliD18, Omand23, Dessart24prep}. From a grid of one-zone models, \citet{Omand23} successfully matched individual spectra of SN~2012au and inferred that a $\sim$15~ms magnetar could be powering it. The later study by \citet{Dessart24prep} created flux spectra from magnetar-powered explosion models and evolved them in time. The magnetar+clump models that successfully matched multiple epochs of SN~2012au observations were those that invoked greater clumping, which reduced the ejecta ionization. The extended loop signature seen at day 26 in our observation of SN~2012au, which we propose to be caused by a hot, helium-absorbing region, may be evidence of such clumping. From their magnetar-powered models, \citet{Dessart24prep} suggested SN~2012au remained infrared luminous for 5--10 years after explosion, while \citet{Omand23} predicted it would remain radio bright ($F>100~\mu$Jy) for decades. Follow-up observations would provide valuable tests
of these model predictions. 

\section{Conclusions}
Over the first 90 days post-$R$ maximum, continuum polarization values for SN~2012au ranged between  $p=1.17\% \pm0.05\%$ at day 26 and $p=0.19\% \pm0.12\%$ at days 85--90, with a general decreasing trend over time. These values suggest the photosphere of SN~2012au deviated from spherical symmetry on the order of $10-40\%$ throughout its photospheric evolution \citep{Hoflich91}. Additionally, during the early photospheric phase (days 0--40), the bulk of the data formed a clear dominant axis in the $q-u$ plane, with a persistent PA of $\theta\approx 16\degr$ (~\S\ref{sec: continuum}).

Line polarization signatures in SN~2012au also exhibited large magnitudes throughout the photospheric phase. All three major He \textsc{i} lines ($\lambda$5876, $\lambda$6678 and $\lambda$7065) had $p\ge0.63\% \pm0.05\%$, with PAs of $96\degr<\theta<151\degr$, while the Fe \textsc{ii} triplet ($\lambda\lambda\lambda$4924, 5018, 5169) had $p\ge1.93\% \pm0.05\%$ with a constant PA of $\theta\approx23\degr$ (Table~\ref{Tab:phot_line_pol}). Polarization signatures for these two elements were particularly significant during the early photospheric phase, as the iron line traced the dominant axis while the helium lines deviated by $\sim90\degr$ from it, creating loops on the $q-u$ plane (Fig.~\ref{fig:He_lines_QU_vel}). 

When comparing these SP signatures to those of other SNe Ib, we found that differences between the PAs of the He \textsc{i} and Fe \textsc{ii} lines are common in most of the sample at early times (\S~\ref{subsec: disc_comp_other_Ibs}). This difference in PA suggests it is common for these two elements to be distributed differently in SN Ib ejecta. However, the contrast between the two is particularly pronounced in SN~2012au, where these line signatures span a greater velocity range than in any of the other SNe Ib studied with SP. We conclude that the lighter helium material is broken up into more localized hotter areas in the ejecta, while the heavier, cooler, iron material is more evenly distributed in a geometry similar to that of the photosphere (\S~\ref{subsec: disc_structure_over_time}). 

Interestingly, the rare late SP observation (day 295) we present here for SN~2012au also exhibits overall high polarization magnitudes and a newly aligned dominant axis ($\theta\approx 169\degr$). This last epoch coincides with the start of the ejecta's slow transition to the nebular phase \citep{MiliD18, Pandey21}. From this we conclude that the decrease in opacity during this transition allowed  us to see a highly asymmetric interior structure, which may retain a direct imprint of the core explosion mechanism. Due to the complex nature of this late observation, we cannot distinguish whether this structure is the deeply-receded photosphere or whether it represents heavy elements specifically distributed $90\degr$ from the PA of the late-time dominant axis. However, we can confidently report that the orientation of the bulk ejecta at day 295 lies at least $60\degr$ from its orientation during the early photospheric phase, indicating multi-layered, dual-axis ejecta. 

We propose that the dual-axis ejecta structure we observe for SN~2012au is a signature of a preferentially directed explosion mechanism,  which is oriented nearly face-on to our line of sight (providing an optimal viewing of the ejecta structure; \S~\ref{subsec: disc_structure_over_time}; Fig.~\ref{fig:pos_geom}). This picture of SN~2012au may explain its higher energy and luminosity than typical CCSNe. The difference in orientation between the components of 2012au's ejecta structure is also reminiscent of the magnetohydrodynamical effects of a non-standard magnetized, rotating star suggested as a magnetar progenitor \citep[][\S~\ref{subsec: disc_mechanism}]{mikami08}, and is consistent with the nebular-phase powering source proposed by \citet{Omand23, Dessart24prep}. 

The polarization signatures observed in SNe Type Ib show they form a range of asymmetric explosions \citepalias{Maund07, Maund09, Tanaka09, Reilly16}. This work suggests that SN~2012au defines the more extreme end of this range, while identifying common traits among the SN Ib SP sample (\S~\ref{subsec: disc_comp_other_Ibs}). The level of continuum polarization for these objects varies, consistent with ellipsoidal electron-scattering photospheres deviating by 10--50\% from spherical symmetry \citep{Hoflich91}. In each case, line polarization behavior deviating from that of the continuum indicates that the photospheric light is partially blocked. Loop signatures are prevalent in all of these SNe, suggesting that even more complex geometries, such as multiple clumps, are common in their ejecta structures. 

Compared with this sample, the polarization behavior of SN~2012au is most similar to that of SN~2005bf. Both these SNe exhibit a dominant axis at early times and higher levels of continuum polarization than the others in the sample, suggesting that they are viewed at a more optimum (face-on) angle to see their elongated structure (\S~\ref{subsec: disc_comp_other_Ibs}). However, the polarization levels seen throughout the evolution of SN~2012au are higher still than SN~2005bf (Table~\ref{Tab:Ib_SNe}), suggesting it is the most asymmetric of the sample; this extreme behavior may be related to its connection with hypernovae, superluminous SNe, and magnetar hosts. Additional multi-epoch SP observations of SNe Type Ib are required to decipher whether the characteristics we infer for SN~2012au are common among this SN type, or further isolate this object as a peculiar piece in the classification puzzle.

\section{Acknowledgments}
Observations reported here were obtained at the MMT Observatory, a joint facility of the University of Arizona and the Smithsonian Institution. Observations were also obtained using Steward Observatory facilities, available to astronomers at the University of Arizona, Arizona State University, and Northern Arizona University. S.D. and J.L.H.  acknowledge support from NSF award AST-2009996. They also recognize that the University of Denver resides on the ancestral territories of the Arapaho, Cheyenne, and Ute nations and that its history is inextricably linked with the violent displacement of these indigenous peoples. D.C.L. acknowledges support from NSF grants AST-1009571, AST-1210311, and AST-2010001, under which part of this research was carried out. 

\appendix

\section{ISP Estimates for SN 2012au}\label{sec: Appendix ISP}

The contribution of local Galactic dust to the total ISP for a supernova can be estimated by observing distant Milky Way stars near the line of sight of the target and assuming all of their measured polarization is due to the Milky Way ISP \citep[e.g.,][]{Tran1995,Stevance17}. Following this method, we compared the polarization of HD 112142, a Galactic M star within $2\degr$ of SN~2012au \citep{Heiles} to the ISP level inferred by \citet{Pandey21} from a fit to 9 probe stars within $10 \degr$ of SN~2012au. We found an overlap in ISP magnitude between HD 112142 ($p = 0.267\% \pm 0.036\%$, $\theta$ = $65.7\degr \pm 3.9\degr$) and the value quoted by \citet{Pandey21} for the best fit to their 9 probe stars ($p = 0.23\% \pm 0.01\%$, $\theta$ = $127.7\degr \pm 1.1\degr$), but the polarization angles of these two ISP estimates do not agree. This could be due to the fact that HD 112142 is a long-period variable star and thus may not be the best candidate for an ISP probe  \citep{Campbell1955}. In order to provide the best sample of the Milky Way ISP, probe stars should be at least at a distance comparable to the thickness of the galactic dust plane at the target's latitude \citep{Tran1995}. For SN 2012au we calculated this to be $d > 187$ pc. We display both the HD 112142 and \citet{Pandey21} probe star ISP estimates in Figure \ref{fig:ISP_est} as the the pink and green crosses, respectively. We also show the polarization value for only the most distant star from the sample used by \citet{Pandey21}, HD 112325 ($d = 870$ pc; \citealt{gaia2018collab}) as the purple cross for comparison. The percent polarization of HD 112325 ($p = 0.22\% \pm 0.05\%$) agrees with other probe star estimates, but its PA is different ($\theta$ = $53.03\degr \pm 6.01\degr$). If the host ISP contribution is directed such that it adds to rather than subtracts from the Milky Way portion, then the polarization values quoted here can be taken as lower limits on the total ISP. However, the host ISP could also be directed such that it completely cancels with the Milky Way polarization, resulting in a lower limit of 0. Each of these scenarios highlights the importance of knowing the PAs of both ISP components in order to know the result of their vector addition. We explore scenarios for various host ISP PAs in Appendix \ref{sec: Appendix Serk.}. 

We obtained a third estimate for the ISP toward SN 2012au by examining the polarization in the continuum regions of our SPOL spectrum from days 85--90 (\S~\ref{sec: continuum}). At later epochs, as the SN ejecta dissipate and become optically thin to electron scattering \citep[e.g.,][]{Porter16}, the continuum polarization level should be largely dominated by the ISP. We chose the spectrum from days 85--90 because our later data were obtained at day 295, well into the transitional-nebular phase of the SN, when few other SP observations have been made and our understanding of the behavior of the continuum polarization is less developed. While one might expect very low SN continuum polarization at that stage, recent models argue this may not always be the case \citep{Dessart24}. Additionally, our day 295 spectrum  does not extend to long enough wavelengths to include the redder (7250--7350~\AA) of the two continuum regions we defined for earlier epochs, and the bluer region (5120--5270~\AA) overlaps with a clear emission line, making it a less suitable choice for a continuum estimate than in previous epochs (Fig.~\ref{fig:specpol}, \S~\ref{sec: continuum}). From our two continuum regions for days 85--90, we calculated a weighted average $q = 0.10\% \pm 0.06, u = -0.17\% \pm 0.06\%$, resulting in $p_{max}=0.19\% \pm 0.06\%$ and $\theta = 150.1\degr \pm 9.7\degr$. This point is shown in Figure~\ref{fig:ISP_est} as the yellow cross in the fourth quadrant.  

Another method we explored to derive the ISP was based on the assumption that strong emission lines are intrinsically unpolarized. While this assumption has been demonstrated to be adequate for SNe Type II \citep{Dessart21b,Leonard21}, we found it challenging to identify good line candidates for this method. \citetalias{Reilly16} used the Ca \textsc{ii} IR triplet present in their day +36 spectra of SN~iPTF 13bvn; however, our spectra do not extend this far into the infrared. Additionally, the use of a single epoch for this derivation does not verify the unchanging aspect of the ISP over time. Instead we chose to investigate the three most prominent He \textsc{i} lines (\S~\ref{subsec: Helium}) in the early photospheric phase spectra (days 0--7, 26 and 35--40). We compared 9 total polarization estimates, from these three line regions over the three epochs, and they all had magnitudes less than or in agreement with that yielded by the Serkowski $E(B-V)$ estimate. However, their PAs were not consistent with each other (22.5\degr~$< \theta <$ 151.8\degr $\pm12.5$\degr). The error-weighted average of these line regions is shown in Figure~\ref{fig:ISP_est} as a black cross.

Our final ISP estimate lies at the intersection of the best-fit lines for the dominant axis in the $q-u$ plane (\S~\ref{sec: continuum} and Fig.~\ref{fig:all_QU}) at days 0--7 (blue) and day 295 (grey); this stems from the assumption that the ISP is located somewhere along the dominant axis of each observation. We chose the first and last epochs because the ISP should remain constant over time, so the intersection of these two lines represents this consistency. From this location in the $q-u$ plane we calculated an ISP estimate of $p = 0.50\% \pm 0.19\%, \theta = 50.03\degr \pm 11.8\degr$. This point is located in the second quadrant of the $q-u$ plane, inside the Serkowski reddening estimate upper limit; its PA is consistent with that of the farthest probe star, HD 112325. 

In Section~\ref{sec: ISP} we present our chosen ISP estimate, which by methodology alleviates the dependencies found between each estimate detailed here.

\section{Serkowski Fit ISP Estimates}\label{sec: Appendix Serk.}

\citet{Serkowski75} introduced a single empirical relationship that could model the expected wavelength dependence of interstellar polarization (ISP) across an optical spectrum, given the observed maximum polarization $p_{max}$ and corresponding wavelength $\lambda_{max}$: 

\begin{equation}
p(\lambda)=p_{max} \textrm{exp}[-K \textrm{ln}^2(\lambda_{max}/\lambda)]
    \label{eq:Serk}
\end{equation}

In more recent studies, the \textit{K} value used is itself wavelength-dependent \citep[e.g.,][]{Wilking82}:

\begin{equation}
K = (-0.1\pm0.05) + (1.86\pm0.09)\lambda_{max}
\end{equation}

The $p(\lambda)$ relationship given in Eq.~\ref{eq:Serk} is generally known as a \textit{Serkowski law}, which hypothetically can be used to infer the wavelength-dependent ISP contribution to many astronomical polarization observations, including those of SNe. However, this model was based on observations of stars in the Milky Way (MW). To accurately account for the effects of ISP in SN~observations, one must consider that the MW and host galaxy of the observed object both contribute to the ISP. The vector sum of these two contributions can result in a range of polarization magnitudes and position angles. Thus, the assumption that the true total ISP can be described by a single Serkowski curve may not always be justified. In this appendix we explore the parameter space created by vectorially adding two theoretical Serkowski curves, one representing the Milky Way and one representing the host galaxy. To do so we assumed the PAs of both curves are constant with wavelength. 

\begin{figure*}
    \centering
    \includegraphics[width =.8\textwidth]{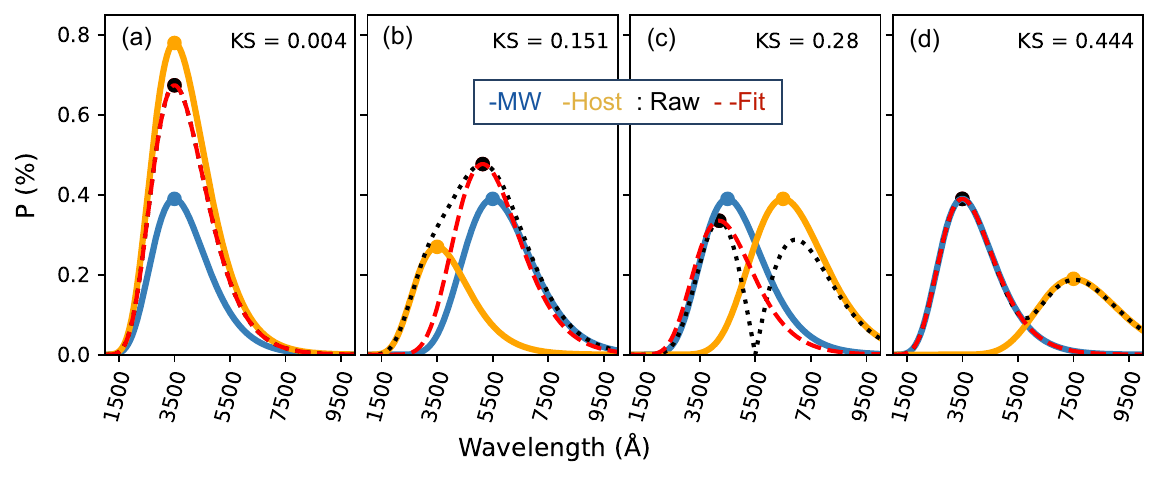}
    \caption{Serkowski curves for a representative sample of the parameters we explored. Each panel contains a Milky Way curve (blue solid line), a host galaxy curve (orange solid line), a ``Raw'' combined curve (MW + Host; black dotted line), and a ``Fit'' curve constructed from the $P_{max}$ and $\lambda_{max}$ values of the Raw curve (red dashed line), as well as the corresponding KS statistic for each set of curves (see Appendix \ref{sec: Appendix Serk.} for details). The peak polarization value for each curve is denoted by a round point in the respective color.}
    \label{fig:Serk_curves}
\end{figure*}

We began this investigation by creating a Serkowski law from a representative polarization value for the region of the Milky Way through which SN~2012au is viewed (this value should of course be modified for full treatment of objects along other sightlines). To determine the magnitude of the MW peak polarization, $p_{max,MW} (\%)$, we used the reddening value for the MW in the direction of SN~2012au discussed in \citet{schlafly2011} and assumed maximum polarization efficiency, resulting in a value of $p_{max,MW} (\%) = 0.39\%$. For the MW position angle along this sightline, we used the PA that \citet{Pandey21}~reported for the most distant of their probe stars, HD 112325 ($d = 870$ pc; \citealt{gaia2018collab}): $\theta_{MW} = 53.03\degr \pm 6.01\degr$. 

Rather than adopting a single value for the MW wavelength of peak polarization, $\lambda_{max,MW}$, we chose to vary this parameter. According to a survey conducted by \citet{Serkowski75}, who  explored the $UBVR$ spectral regions of 180 (mostly southern) stars, on average the MW ISP curve peaks at $\lambda_{max,MW}= 5500$ \AA~. However, these authors noted that $\lambda_{max}$ varies throughout the optical wavelengths\footnote{\citet{Wilking80, Wilking82} investigated the shift of the Serkowski curve when $\lambda_{max}$ lies in the near-IR or near-UV; however, because our SNSPOL observations are optical only, consideration of this shift is beyond the scope of this work.} for different MW sightlines. Additionally, evidence of the ISP peaking at a shorter wavelength was reported by \citetalias{Tanaka09}. To account for this, we also tested four other values of $\lambda_{max,MW}$ spanning the optical region (Table \ref{tab:Serkowski_parms}). We display the Serkowski curves for a sample of these MW parameters as the blue curves in Figure \ref{fig:Serk_curves}.

To investigate the scenario in which the dust in the host galaxy also follows a Serkowski law, we constructed a separate Serkowski curve to represent the host contribution. However, the parameters from which to construct this Serkowski curve are much less well known, creating the majority of the parameter space explored in this study (Table \ref{tab:Serkowski_parms}). For the host galaxy peak wavelength, $\lambda_{max,H}$, we used the same sample of wavelengths as for $\lambda_{max,MW}$. Then to characterize the peak polarization of the host, we defined $p_{max,H}$ values in relation to the reasonably well determined MW value. We tested several cases: (I) $p_{max,H}$ being (roughly) half the magnitude of $p_{max,MW}$ (\S~\ref{sec: ISP}); (II) the host ISP obeying the Serkowski reddening relationship with $E(B-V)=0.03$ \citep{MiliD13} so that the maximum combined host + MW polarization equals our estimated ISP upper limit (\S~\ref{sec: ISP}); (III) the rare case of $p_{max,H} = p_{max,MW}$; (IV) $p_{max,H} = 2p_{max,MW}$; and (V) the host dominating the total ISP, with $p_{max,H} = 5p_{max,MW}$. The resulting set of $p_{max}$ values probes roughly the same range of total ISP values found in the other SNe Ib spectropolarimetry studies (\citetalias{Maund07}, \citetalias{Maund09}, \citetalias{Tanaka12} and \citetalias{Reilly16}): a few tenths of a percent to a few percent (Table~\ref{tab:Serkowski_parms}). 

We chose PA values for the host component of the ISP ($\theta_{H}$, which remained constant with wavelength) such that the sample covered at least one angle from each of the quadrants in the $q-u$ plane. Among the sampled angles we made sure to include the case in which $\theta_{H}=\theta_{MW}$ so that the two components  perfectly add, and the case in which $\theta_{H}=\theta_{MW}-90\degr$
so the two polarization components perfectly subtract. We also chose $\theta_{H}=172.93\degr$ for one of our cases, so that $\theta_{H}+\theta_{MW}$ combine to equal the angle of the intersection of the best fits to our earliest and latest epochs of SN~2012au data (days 0--7 and 295; Fig.~\ref{fig:ISP_est}, \S~\ref{sec: ISP}). 

\begin{table}
 \centering
 \begin{tabular}{|c|c c c c c|}
    \hline
    $\lambda_{max,MW}$ (\AA) & 3500 & 4500 & 5500 & 6500 & 7500\\
    \hline
    $\lambda_{max,H}$ (\AA) & 3500 & 4500 & 5500 & 6500 & 7500\\
    \hline
    $p_{max,H} (\%)$ & 0.19 & 0.27 & 0.39 & 0.78 & 1.95\\
    \hline
    $\theta_{H} (\degr$) & 0 & 53.03 & 106.06 & 143.03 & 172.93\\
    \hline
 \end{tabular}
 \caption{Parameters for the Milky Way ($MW$) and host galaxy ($H$) Serkowski laws we combined. Varying each of those values with respect to the others creates a space of 625 unique test cases for the ISP conditions. For the Milky Way ISP component, $p_{max,MW} = 0.39\%$ and $\theta_{MW} = 53.03\degr$ were kept constant. }
 \label{tab:Serkowski_parms}
 \end{table}

Varying each of the values in Table \ref{tab:Serkowski_parms} with respect to the others created a parameter space of 625 unique test cases for the total ISP conditions. For each combination of MW and host galaxy parameters, our analysis began by creating individual Serkowski curves from Equation \ref{eq:Serk}. These Serkowski curves are plotted as the blue and orange curves in Figure \ref{fig:Serk_curves}, for the MW and host, respectively. We then decomposed both the MW and host galaxy Serkowski curves into their respective \textit{q} and \textit{u} components in order to calculate the new total \textit{p} function by adding the two together in quadrature. We refer to the straightforward addition of the two curves, with no other assumptions imposed, as the 
``Raw'' combined spectrum, shown as the black-dotted curves in Figure \ref{fig:Serk_curves}. Often this Raw spectrum did not resemble a Serkowski curve, but instead appeared to be broader or double-peaked. With this in mind, we then took steps to quantify the discrepancies. 

To do so we constructed a new Serkowski spectrum from the maximum \textit{\%p} in the Raw combined curve and its corresponding wavelength; we called this a ``Fit'' spectrum. These 
``Fit'' curves are displayed as the red-dashed curves in Figure \ref{fig:Serk_curves}. We then quantified the difference between the 
Raw and Fit curves by calculating the Kolmogorov-Smirnov (KS) goodness-of-fit statistic, using the 
Fit spectrum as the expected function \citep{Hodges58}. Applying the \textit{two-sided} prescription, the null hypothesis is that the distributions of the two functions are identical, i.e., $F(x)=G(x)$ for all $x$. For our application, the KS statistic is then the maximum absolute difference between the empirical distribution of the Raw and Fit curves. The KS statistic ranges from $0-1$; the greater the difference between the Raw and Fit
curves, the closer to 1 the KS statistic.

\begin{figure}
    \centering
    \includegraphics[width =\columnwidth]{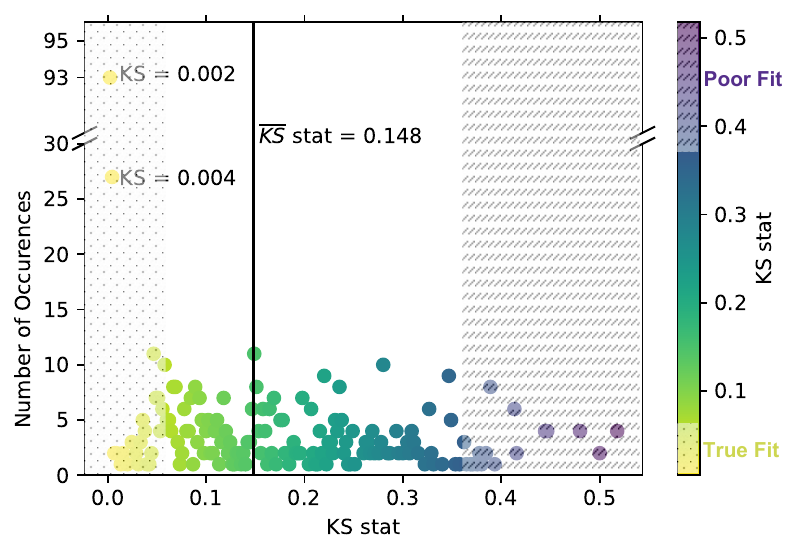}
    \caption{Occurrences of values of the KS statistic for the tested parameter space. Regions corresponding to the ``true fit'' and ``poor fit'' categories (based on the KS statistic; Appendix \ref{sec: Appendix Serk.}) are filled with gray dots and hashes, respectively. The categories ``good fit'' and ``questionable fit'' both span the middle unfilled region. The average KS statistic is shown as a black vertical line with the value labeled.}
    \label{fig:Serk_KS_occurences}
\end{figure}

For our parameter space of 625 combined MW + Host Serkowski curve cases, the spread of the resulting KS statistics and number of occurrences of each in the tested space are shown in Figure \ref{fig:Serk_KS_occurences}. In Figure \ref{fig:Serk_curves}, we display a set of curves that 
represent the variety of shapes among the Raw functions in our sample,
along with the respective KS statistic for each scenario. 
The KS statistics for our full sample range from $0.002-0.518$. Within this range of KS statistics, we determined four categories to describe the extent to which the Fit functions were an accurate representation of their corresponding Raw curves: ``true fits'', ``good fits'', ``questionable fits'' and ``poor fits'' (Fig.~\ref{fig:Serk_KS_occurences}). We describe these categories and their criteria below. 

The most occurrences of a single KS statistic are for the lowest value, KS = 0.002, for which the Raw and Fit functions are indistinguishable. Furthermore, we found that the two functions remain visually indistinguishable for KS \textless~0.053, defining our ``true fits'' category. Of the 625 cases we tested, 28\% fall into this category (Fig. \ref{fig:Serk_curves}, panel (a)). Because the Fit curves each have a single peak by construction, all the Raw functions in the ``true fit'' category are also single-peaked. On the other end of the scale, we find that all Raw functions with KS \textgreater~0.364 are double-peaked; we categorize these as ``poor fits'' (panel (d) of Fig. \ref{fig:Serk_curves}). Only 7\% of the total cases fall into this category. 

\begin{figure}
    \centering
    \includegraphics[width =\columnwidth]{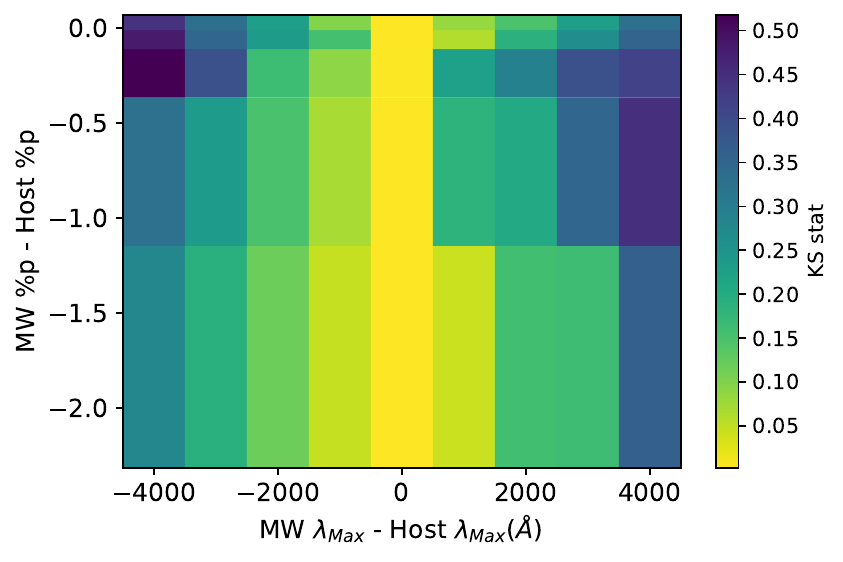}
    \caption{KS statistic mapped by the difference in $p_{max}$ and $\lambda_{max}$ values between the MW and host galaxy.}
    \label{fig:Serk_KS_P}
\end{figure}

\begin{figure}
    \centering
    \includegraphics[width =\columnwidth]{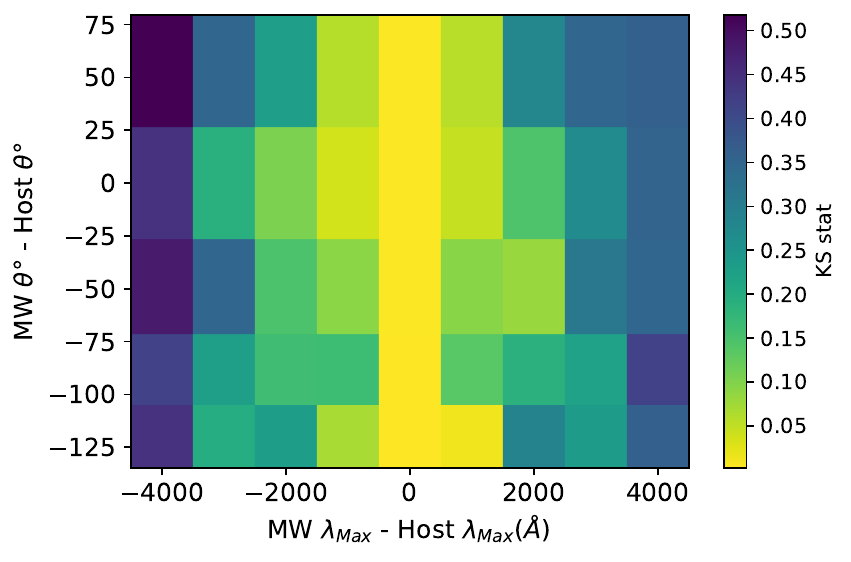}
    \caption{KS statistic mapped by the difference in $\theta$ and $\lambda_{max}$ values between the MW and host galaxy.}
    \label{fig:Serk_KS_PA}
\end{figure}

Figures \ref{fig:Serk_KS_P} and \ref{fig:Serk_KS_PA} depict the behavior of the KS statistic as a function of the MW--host differences in the three Serkowski parameters. The distinct double-peaked shape of the Raw combined functions in the ``poor fits'' category is a result of $\lambda_{max,MW}$ and $\lambda_{max,H}$ being separated by 3000 \AA~or more. Conversely, all the Raw functions in the ``true fits'' category have $\lambda_{max,MW}-\lambda_{max,H} \le 1000$ \AA. This difference in peak wavelength is the primary influence on the KS statistic in these two categories; Figures \ref{fig:Serk_KS_P} and \ref{fig:Serk_KS_PA} show that for low and high KS values, the differences between the MW and host $p_{max}$ and $\theta$ values do not strongly correlate with the KS statistic. 

A KS value between the ``true fit'' upper limit and the ``poor fit'' lower limit ($0.053 \leq$ KS $\leq 0.364$) may describe either a single- or a double-peaked Raw combined curve (panels (b) and (c) of Figure \ref{fig:Serk_curves}, respectively). Within this range, we categorized single-peak Raw curves as ``good fits'' and double-peaked Raw curves as ``questionable fits''; these categories comprise 31\% and 33\% of our total test cases, respectively. For these intermediate KS categories, we found that in addition to the separation between $\lambda_{max,MW}$ and $\lambda_{max,H}$, the difference between $\theta_{MW}$ and $\theta_{H}$ also had a strong influence on whether the Raw combined curves were single- or double-peaked. For the ``good'' (single-peaked) fits, the maximum value of the peak wavelength separation is 3000 \AA, and 92\% of the ``good'' cases have peak wavelength separations of either 1000 \AA~or 2000 \AA. Conversely, the peak wavelength separation for the ``questionable fits'' spans all possible values (0--4000 \AA), but 70\% of these cases have separations of either 2000 \AA~or 3000 \AA. Additionally, only 1.4\% of the ``good'' cases have a $90\degr$ difference between $\theta_{MW}$ and $\theta_{H}$, while 50\% of the ``questionable'' cases do. 

There are 5 cases not captured by one of these four categories, which are those with $P_{max,MW}$ = $P_{max,H}$, $\lambda_{max,MW}$ = $\lambda_{max,H}$, and a $90\degr$ difference between $\theta_{MW}$ and $\theta_{H}$. This occurred because the calculation of the Raw function 
(and subsequent Fit curve) 
results in $F(x) = G(x) = 0$ due to the cancellation of the orthogonal Stokes vectors. However, the KS statistic for these 5 cases ranges from 0.157 to 0.240 for $\lambda_{max,MW} = \lambda_{max,H} =$ 3500 \AA~to 7500 \AA, respectively. This suggests that the calculated KS statistic may be slightly skewed towards shorter $\lambda_{max}$ values. 

We conclude that under the assumption that the dust in both the MW (in the direction of SN~2012au) and the host galaxy (NGC 4790) follows the Serkowski relation, the combination of these two Serkowski curves results in a function closely resembling a Serkowski curve (``true fits'' and ``good fits'') only about half of the time. The rest of the time, the combination of these two Serkowski curves results in a double-peaked function (``questionable fits'' and ``poor fits'') that does not closely resemble a Serkowski curve. We find  that the difference between $\lambda_{max,MW}$ and $\lambda_{max,H}$, has the most influence on whether the Raw combined function resembles a Serkowski (Fit) curve, according to our calculated KS statistic value. This highlights the importance of knowing the polarization peak wavelength for each of the individual components. Additionally, we found that the difference between $\theta_{MW}$ and $\theta_{H}$ has a substantial influence on whether the resulting Raw combined function has a single- or double-peaked shape for half of the tested cases, demonstrating that knowledge of the contributing components' PA values is also important. 

We further conclude that if the difference in peak polarization wavelength between the MW and host galaxy are known to be less than 1000 \AA~apart, then the combined effects of the dust in both galaxies may be assumed to behave like a Serkowski law. In these cases, one can use the single polarization peak location to derive a wavelength-dependent ISP estimate, regardless of whether the position angles of each component are known. However, if the difference in peak polarization wavelength is greater than 1000 \AA, it is more likely that the combined effect of the dust in both galaxies does not behave like a Serkowski law. Under such circumstances, the behavior of the dust may result in multiple polarization peaks. Thus in these cases, using a Serkowski law to derive a wavelength-dependent ISP estimate is not a safe assumption, unless one also knows the PA of the ISP contribution from each galaxy. Fortunately, if one only knows the peak polarization magnitude of the each component, the straightforward addition of the two (without assuming any condition on the PA of each, as opposed to the vector sum of the two) can be used to set an ISP upper limit across all wavelengths, as we have done in our analysis of SN~2012au (\S~\ref{sec: ISP}). This is shown by the fact that even in the worst ``poor fit'' cases (when there are two distinct polarization peaks in the Raw combined curve), the magnitude of either peak must be less than or equal to $p_{max,MW}+p_{max,H}$. 

Finally, we reiterate that these results are based on this specific case study with a parameter space defined by our target of interest, SN 2012au. We acknowledge that to more fully understand the functions that can result from adding two Serkowski curves, a larger parameter space should be investigated.  


\label{lastpage}
\bibliography{SN2012au}{}
\bibliographystyle{aasjournal}




\end{document}